\documentclass[%
preprint,
 amsmath,amssymb,
 aps,
]{revtex4-1}

\usepackage{bm}
\usepackage{graphicx}
\usepackage{epstopdf}
\usepackage{subfigure}
\usepackage{color}

\begin{document}

\title{Phase ordering kinetics of second-phase formation near an edge dislocation}

\author{C. Bjerk\'en and A. R. Massih}
 \altaffiliation[Also at ]{Quantum Technologies AB, Uppsala Science Park, SE-751 83 Uppsala, Sweden}
\affiliation{%
Division of Materials Science, School of Technology, Malm\"{o} University, SE-205 06 Malm\"{o}, Sweden\\
}%




\date{\today}

\begin{abstract}

The time-dependent Ginzburg-Landau (TDGL) equation for a single component non-conservative structural order parameter is used to study the spatio-temporal evolution of a second phase in the vicinity of an edge dislocation in an elastic crystalline solid. A symmetric Landau potential of sixth-order is employed. Dislocation field and elasticity modify the second-order and fourth-order coefficients of the Landau polynomial, respectively, where the former makes the coefficient singular at the origin.  The TDGL equation is solved numerically using a finite volume method, where a wide range of parameter sets is explored. Computations are made for temperatures both above and below the transition temperature of a defect-free crystal $T_{c0}$. In both cases, the effects of the elastic properties of the solid and the strength of interaction between the order parameter and the elastic displacement field are examined. If the system is quenched below $T_{c0}$, steady state is first reached on the compressive side of the dislocation. On the tensile side, the growth is held back.  The effect of thermal noise term in the TDGL equation is studied. We find that if the dislocation is introduced above $T_{c0}$, thermal noise supports the nucleation of the second phase, and steady state will be attained earlier than if the thermal noise were absent. For a dislocation-free solid, we have compared our numerical computations for a mean-field (spatially averaged) order parameter versus time with the late time growth of the ensemble-averaged order parameter, calculated analytically, and find that both results follow late time logistic curves.

\vspace{1 cm}

\end{abstract}

\maketitle


\section{Introduction}
\label{sec:Intro}

Defect enhanced nucleation of a new phase has been observed and modeled in a variety of crystalline materials through the years \cite{Wilsdorf_Kuhlmann_1954,Hornbogen_1962,Gomez_Ramirez_1973,Schmidth_Schabl_1977,Boulbitch_Toledano_1998,Leonard_Desai_1998,Hin_et_al_2008a}.
In metallurgical systems, for example in aluminium-zinc-magnesium alloys, dislocations induce and enhance nucleation and growth of the coherent Laves phase MgZn$_2$ \cite{Allen_Vandesande_1978,Deschamps_et_al_1999,Deschamps_Brechet_1999}. In ammonium bromide (NH$_4$Br) a new phase appears in the proximity of crystal dislocations \cite{Belousov_Volf_1980}. In titanium and zirconium alloys, containing hydrogen, when the hydrogen content exceeds the terminal solid solubility at a given temperature, hydrides (TiH$_x$, ZrH$_x$) form near dislocations and on grain boundaries, affecting mechanical properties of the alloys \cite{Chen_Li_Lu_2004,Northwood_Kosasih_1983}. In magnetic systems such as gadolinium iron garnet Gd$_3$Fe$_5$O$_{12}$, dislocations not only affect the spin reorientation transition lines in the associating phase diagram, they also change the nature of phase transition \cite{Vlasko_Vlasov_1981,Vlasko_Vlasov_1983}.

Other examples on the effect of dislocations include quantum crystals, metals and semiconductors, in which electrons get localized near an edge dislocation producing discrete levels in certain part of the energy spectrum locally, hence altering the electronic structure of the solid, thereby affecting its thermodynamics and transport properties \cite{Lifshitz_Pushkarov_1970,Natsik_Potemina_1980}. In a superconductor and a Bose solid, Ginzburg-Landau type model \cite{Annettt_2004} calculations show that superconductive and superfluid phase nucleate first (at higher temperatures) on dislocations  prior to the bulk of the crystal \cite{Nabutovskii_Shapiro_1978,Goswami_2011,Arpornthip_2011}.

A generic theoretical approach to model the aforementioned phenomena is the Landau-Ginzburg theory of phase ordering. In this paper we use this theory to describe the kinetics of second-phase nucleation in the vicinity of an edge dislocation in solid crystals. A scalar non-conserved order parameter characterizing the presence or the absence of the new nucleus is assumed. The elastic behavior of the solid is taken into account by including the so-called \emph{striction} term in the system free energy, which considers the interaction between the order parameter and deformation \cite{Larkin_Pitkin_1969,Imry_1974}. Moreover, the effect of a dislocation is incorporated in the system free energy through the equation of mechanical equilibrium in the presence of an isolated edge dislocation. The time-dependent Ginzburg-Landau (TDGL) equation with a symmetric tricritical potential energy is solved numerically to evaluate the spatial-temporal behavior of the order parameter in the vicinity of an edge dislocation for different sets of the coefficients in the potential energy. Our model is applicable to systems where second-phase ordering occurs with a preferred orientation of nuclei under an external force. These comprise $\alpha^{\prime\prime}$-phase formation in Fe-N alloys \cite{Sauthoff_1981}, $\theta^{\prime}$-phase nucleation in Al-Cu alloys \cite{Eto_1978,Skrotzki_1996}, hydride formation in titanium \cite{Louthan_1963} and zirconium alloys \cite{Hardie_Shanahan_1975,Massih_Jernkvist_2009}.

We should, though, note that here only the ordering (orientation) aspect of the issue is analyzed. This is characterized by a non-conserved order parameter field variable. The effect of the composition field governed by conserved kinetic equation is decoupled from the TDGL model and is not treated here. A more general set-up with coupled conserved and non-conserved field variables with defects in an elastic solid was presented in \cite{Massih_2011a}. The ground state exact solution with a mean field treatment of the static non-conserved field regarding second-phase nucleation on an edge dislocation was given in \cite{Massih_2011b}.

The organization of this paper is as follows. In Sec. \ref{sec:Model}, we lay down the formalism of the employed model, i.e. the governing TDGL equation and the parameters entering it for the problem under consideration. The numerical method utilized to solve the governing equation is outlined in Sec. \ref{sec:Numeric}. The results of our numerical computations regarding the spatial-temporal evolution of the order parameter in the vicinity of an edge dislocation,  for different values of the phenomenological coefficients in the TDLG equation, are presented in Sec. \ref{sec:Results}. The steady-state solutions and the effect of the selected boundary conditions are also discussed in this section. The influence of the background thermal noise on the solutions is evaluated and discussed in Sec. \ref{sec:Noise}. Section \ref{sec:analytic} presents an analytic computation for late time evolution of the order parameter using Mazenko's approach. In Sec. \ref{sec:Conclude}, we give a summary of our main results, state our conclusions and also make some further remarks concerning generalizations of this work and applications to the real-life second-phase precipitation.


\section{Model description}
\label{sec:Model}

We describe the kinetics of second-phase nucleation on an edge dislocation in a crystalline elastic solid by a time-dependent Ginzburg-Landau theory. A scalar order parameter field $\eta(\boldsymbol{\rho},t)$ depending on space $\boldsymbol{\rho}$ and time $t$ defines the symmetry of the structure and distinguishes the two prevailing phases, the solid solution and the precipitated second phase. Thus, $\eta=0$ describes the high-temperature solid solution and $\eta\neq0$ the second-phase nucleus. The non-conserved order parameter $\eta(\boldsymbol{\rho},t)$ obeys the TDGL or Ginzburg-Landau-Khalatnikov (GLK) equation  with a thermal noise $\vartheta_{a}$ \cite{Landau_Khalatnikov_1954,Lifshitz_Pitaevskii_1981}, which corresponds to model A in the classification scheme of  Hohenberg and Halperin \cite{Hohenberg_Halperin_1977}, viz.
\begin{equation}
\frac{\partial\eta}{\partial t}=-\Gamma_{a}\frac{\delta\mathcal{F}}{\delta\eta}+\vartheta_{a}(\boldsymbol{\rho},t).\label{eqn:glk-eq}
\end{equation}
Here, $\Gamma_{a}$ is a kinetic coefficient that characterizes the interface boundary mobility and $\mathcal{F}$ is the total free energy of the system expressed by \cite{Imry_1974,Ohta_1990,Massih_2011b}
\begin{eqnarray}
\mathcal{F} & = & \int\Big[\frac{g_{a}}{2}(\nabla\eta)^{2}+\mathcal{V}(\eta)+\alpha\nabla\cdot\mathbf{u}\,\eta^2\nonumber \\
&  & +\Big(\frac{K}{2}-\frac{M}{d}\Big)\big(\nabla\cdot\mathbf{u}\big)^{2}+M\nabla\mathbf{u}\textbf{:}\nabla\mathbf{u}\Big]d\boldsymbol{\rho}.
\label{eqn:fe-tot}
\end{eqnarray}
Here $g_{a}(\nabla\eta)^{2}$ arises from the inhomogeneity of $\eta$ and is related to short-range interactions, with $g_{a}$ taken as a positive constant. This term accounts for the existence of interfaces within an equilibrium inhomogeneous system \cite{Desai_Kapral_2009}. The second term in the integrand is the Landau potential \cite{Landau_Lifshitz_1980} in the form
\begin{eqnarray}
\mathcal{V}(\eta)=\frac{1}{2}r_{0}\eta^{2}+\frac{1}{4}u_{0}\eta^{4}+\frac{1}{6}v_{0}\eta^{6},\label{eqn:landau-pot}
\end{eqnarray}
where the coefficients $r_{0}$ and $u_{0}$ are temperature-dependent variables while the term $v_{0}\eta^{6}$ with $v_{0}>0$ is needed for the stability when $u_{0}<0$. The term $\alpha\eta^{2}\nabla\cdot\mathbf{u}$ in Eq. (\ref{eqn:fe-tot}) describes the interaction between the order
parameter and the displacement vector field $\mathbf{u}$, where the strength of the interaction is designated by $\alpha$ and is assumed to be a positive constant. The second line in Eq. (\ref{eqn:fe-tot}) is the elastic free energy, where $K$ and $M$ are the bulk and shear modulus, respectively; $d$ is the space dimensionality; $\nabla\mathbf{u}=(\partial u_{j}/\partial x_{i}+\partial u_{i}/\partial x_{j})/2$ is the strain tensor, and the symbol $\textbf{:}$ denotes the tensorial product contracted on two indices. The space integral in Eq.~(\ref{eqn:fe-tot}) is over the volume of the system.

The function $\vartheta(\boldsymbol{\rho},t)$ in Eq. (\ref{eqn:glk-eq}) stands for the presence of background random thermal motion, the so-called Langevin noise, prevailing at temperature $T$ and satisfying the fluctuation-dissipation condition
\begin{eqnarray}
\langle\vartheta_{a}(\boldsymbol{\rho},t)\vartheta(\boldsymbol{\rho}\rq{},t\rq{})\rangle & = & 2k_{B}T\Gamma_{a}\delta(\boldsymbol{\rho}-\boldsymbol{\rho}\rq{})\delta(t-t\rq{}),\\
\langle\vartheta_{a}(\boldsymbol{\rho},t)\rangle & = & 0
\label{eq:langevin}
\end{eqnarray}
where the averages $\langle\dots\rangle$ are over a Gaussian distribution function representing a Gaussian white noise, and $k_{B}$ is the Boltzmann constant. In this context, one quantity of interest in our study is the temporal evolution of the ensemble average of the square of the local order parameter, namely
\begin{equation}
\mathcal{S}(t)=\langle\eta^{2}(\boldsymbol{\rho},t)\rangle,
\label{eqn:meansquare}
\end{equation}
where the average is taken over random initial conditions (noise) described by a Gaussian distribution function.

\begin{figure}[htbp]
\includegraphics{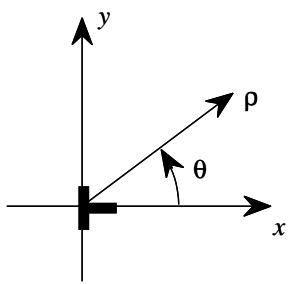}
\caption{Geometry of the edge dislocation.}
\label{fig:GeomDisl}
\end{figure}

Here, we assume that mechanical equilibrium for the displacement $\delta\mathcal{F}/\delta\mathbf{u}=0$ is satisfied at all times. A dislocation generates  local strains that change the equilibrium condition in the solid.  The mechanical equilibrium equation that includes the force field  generated by an edge dislocation in the $xy$-plane with the Burgers vector in the $x$ direction $\mathbf{b}=b\,\mathbf{e}_x$ is
\begin{eqnarray}
M\nabla^{2}\mathbf{u}+(\Lambda-M)\nabla\nabla\cdot\mathbf{u}+\alpha\nabla\eta^{2}=-M b\,\mathbf{e}_y\delta(x)\delta(y)
\label{eqn:mecheq}
\end{eqnarray}
where $\Lambda=K+2M(1-1/d)$,  $\mathbf{e}_y$ is the unit vector along the $y$-axis and $\delta(\bullet)$ is the Dirac delta \cite{Landau_Lifshitz_1970,Massih_2011b}. Equation (\ref{eqn:mecheq}) is then used to eliminate the elastic field from the expression for the total free energy (e.g. \cite{Massih_2011b}), which now can be expressed as
\begin{equation}
\mathcal{F}[\eta]=\int\Big[\frac{g_{a}}{2}(\nabla\eta)^{2}+\frac{1}{2}r_{1}\eta^{2}+\frac{1}{4}u_{1}\eta^{4}+\frac{1}{6}v_{0}\eta^{6}\,\Big]d\boldsymbol{\rho}.
\label{eqn:toten1}
\end{equation}
The last three terms in the integrand correspond to the Landau potential given by Eq. (3) but with altered coefficients for the quadratic and quartic terms to account for the presence of an edge dislocation and an elastic body, respectively:
\begin{eqnarray}
r_{1} & = & \left|r_{0}\right|\big(\textrm{sgn}(r_{0})-\rho_{0}\cos\theta/\rho\big),
\label{eqn:r1-disl}\\
u_{1} & = & u_{0}-2\alpha^{2}/\Lambda,
\label{eqn:u1-elastic}
\end{eqnarray}
where $\rho_{0}\equiv2\alpha bM/(\pi\left|r_{0}\right|\Lambda)$ and $b$ is the magnitude of the Burgers vector. Figure \ref{fig:GeomDisl} shows the geometry of the dislocation. The parameter $\rho_{0}$ can be considered as a local characteristic length related to the presence of the defect in elastic body. For a defect free crystal, $r_{1}=r_{0}$, and for a rigid crystal, $u_{1}=u_{0}$. In more detail, we may rewrite Eq.~(\ref{eqn:r1-disl}) in the form
\begin{eqnarray}
r_{1} & = & a[T-T_{c}(\rho,\theta)],\label{eqn:r1-disl-2}\\
T_{c}(\rho,\theta) & \equiv & T_{c0}+\frac{2\alpha bM}{a\pi\Lambda}\frac{\cos\theta}{\rho}.
\label{eqn:tc-disl-2}
\end{eqnarray}
Here $a$ is taken to be a positive constant, $T_{c0}$ the phase transition temperature in a defect free crystal and $T_{c}(\rho,\theta)$ the phase transition temperature for a crystal with an edge dislocation. The governing  equation for the space-time variation of the order parameter is now obtained by inserting Eq. (\ref{eqn:toten1}) into Eq. (\ref{eqn:glk-eq}):
\begin{equation}
\frac{1}{\Gamma_{a}}\frac{\partial\eta}{\partial t}=g_{a}\nabla^{2}\eta-\big(r_{1}\eta+u_{1}\eta^{3}+v_{0}\eta^{5}\big)+\vartheta_{a}(\boldsymbol{\rho},t).\label{eqn:tdgl-eq2}
\end{equation}
In this study, computations of phase transformation are performed with different combinations of $r_{0}$ and $u_{1}$ by solving Eq.~(\ref{eqn:tdgl-eq2}).


\section{Numerical method}
\label{sec:Numeric}

We have used the open-source partial differential equation solver package \texttt{FiPy} \cite{Guyer_et_al_2009} for our numerical computations. \texttt{FiPy} utilizes a standard finite volume approach, which is extensively used in computational fluid dynamics, in order to reduce the model equations to a form tractable to linear solvers . The spatio-temporal evolution of $\eta$ in a two-dimensional space is computed, which is adequate, since a straight edge dislocation is considered here. The dislocation is placed at the origin, $(x,y)=(0,0)$, with the slip direction along the line $x=0$, see Fig. \ref{fig:GeomDisl}. We in general use a square mesh consisting of $200\times200$ equally-sized square elements, otherwise it is specified. Each element has a side length $\Delta l=\rho_{0}/20$. The gradient of $\eta$ perpendicular to the outer boundaries of the mesh is set equal to zero, i.e. $\textbf{n}\cdot\nabla\eta=0$, where $\textbf{n}$
is a unit vector perpendicular to a boundary, and thus periodic boundary conditions are achieved. In the cases where no thermal noise is included, the initial value of the order parameter, $\eta_{init}$, is taken to be a small positive random number of the order of $10^{-2}$ of the maximal value of $\eta$ obtained in the computations. $\eta_{init}$ is a uniform rectangular distribution varying between $0.005$ and $0.01$. The system is considered to be large enough so that the periodic boundary condition would not affect the results. A reference time increment is defined as $\Delta t_{ref}=0.9\Delta l^{2}/(2g_{a}\Gamma_{a})$. This time step is chosen to be sufficiently small to provide stable solutions for $\eta$ for all the different combinations of $r_{0}$ and $u_{1}$ studied here.


\section{Results }
\label{sec:Results}

The spatio-temporal evolution of $\eta$ is presented for different sets of input parameters: $\{r_0,u_1,v_0,g_a\}$. First the situation when $r_{0}>0$, i.e. $T>T_{c0}$, is investigated; for which, we set $r_{0}=1$. The elastic interaction energy, embedded in $u_{1}$, is varied. Only results for cases with negative values of $u_{1}$ are presented here; $u_{1}=-1,-2,-3$, respectively. This choice will be discussed in this section. The value of the coefficient of the Laplacian term in Eq.~(\ref{eqn:tdgl-eq2}) is set as $g_{a}=0.1$, implying that the influence of the gradient of $\eta$ is relatively large. The influence of $g_{a}$ will be also be discussed here. Thereafter, the behavior for $r_{0}<0$ is studied, namely, $r_{0}=-1$ with $u_{1}=\{-1,1\}$. In all the calculations, $v_{0}$ is put equal to unity, otherwise it is specified. It should be emphasized that the purpose of this study is to explore the characteristics of the ordering evolution in the parameter space $\{r_{0},u_{1}\}\in\mathbf{R}$.


\subsection{Spatio-temporal evolution}
\label{sec:SpatioTempEvol}

\begin{figure}[htbp]
\subfigure[\hspace{1mm} $ t=600\Delta t _{ref} $]{ \label{fig:SurfRpos1Uneg1Time600}
\includegraphics{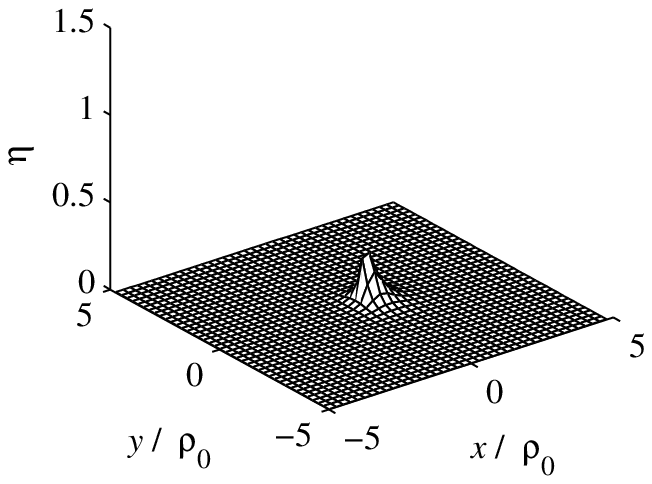}
}
\subfigure[\hspace{1mm} $ t=800\Delta t _{ref} $]{\label{fig:SurfRpos1Uneg1Time800} \includegraphics{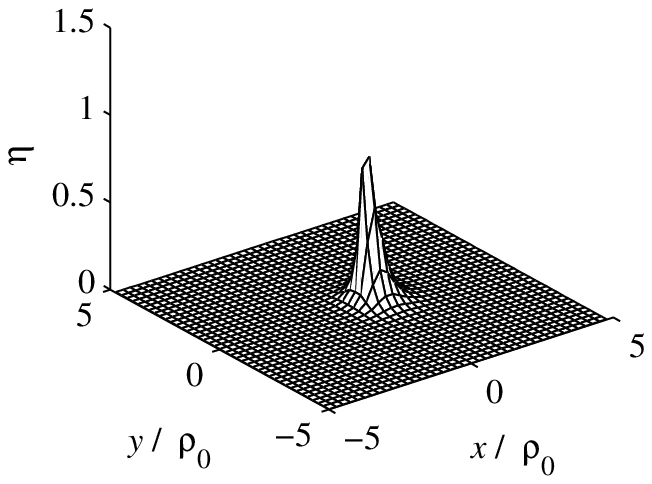}
}
\subfigure[\hspace{1mm} $ t = 1000\Delta t _{ref} $]{ \label{fig:SurfRpos1Uneg1Time1000}
\includegraphics{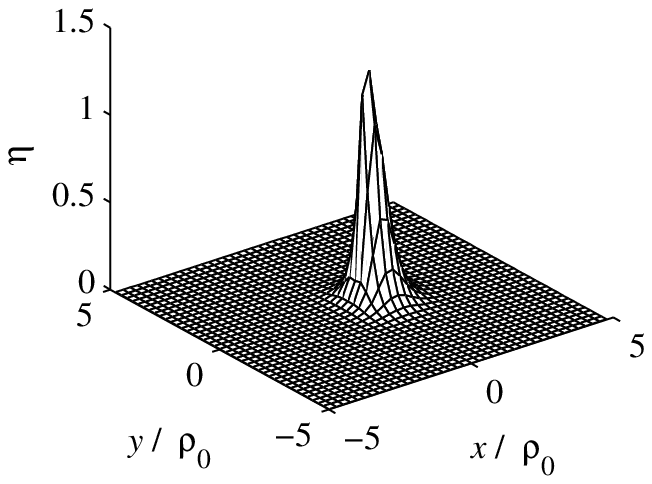} }
\subfigure[\hspace{1mm} $ t=2000\Delta t _{ref} $]{\label{fig:SurfRpos1Uneg1Time2000} \includegraphics{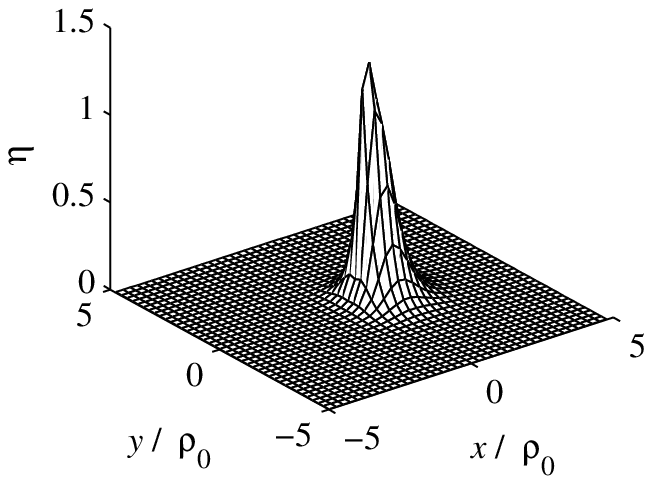}
}
\caption{Spatial distribution of $\eta$ at various times $t=600,800,1000\text{ and }2000~\Delta t_{ref}$,
for $r_{0}=1$, $u_{1}=-1$ and $v_{0}=1$. $\Delta t_{ref}$ is the reference time-step used in computations and is defined in Sec. \ref{sec:Numeric} of the main text. Only every fifth node in the mesh is selected for the illustration.}
\label{fig:SurfsRpos1Uneg1}
\end{figure}

\begin{figure}[htbp]
\centering{}
\subfigure[\hspace{1mm} $ r_0=1$ and $u_1=-1$]{
\includegraphics[width=7cm]{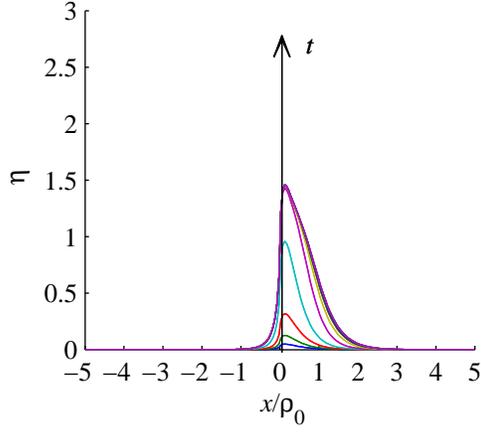}
\label{fig:ProfileY0Rpos1Uneg1}
}
\subfigure[\hspace{1mm} $ r_0=1$ and $u_1=-2$]{
\includegraphics[width=7cm]{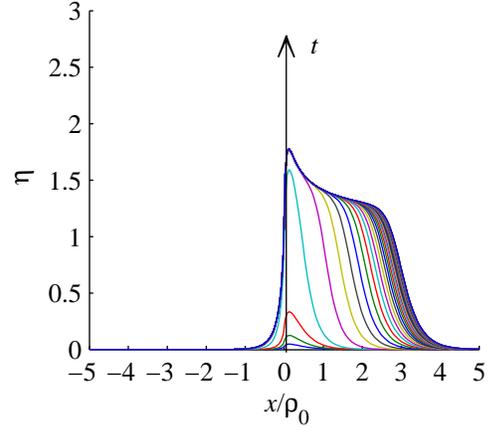}
\label{fig:ProfileY0Rpos1Uneg2}
} \\
\subfigure[\hspace{1mm} $ r_0=1$ and $u_1=-3$]{
\includegraphics[width=7cm]{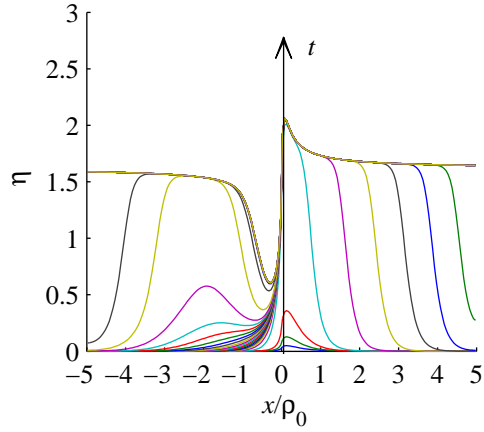}
\label{fig:ProfileY0Rpos1Uneg3}
} \\
\subfigure[\hspace{1mm} $ r_0=-1$ and $u_1=1$]{
\includegraphics[width=7cm]{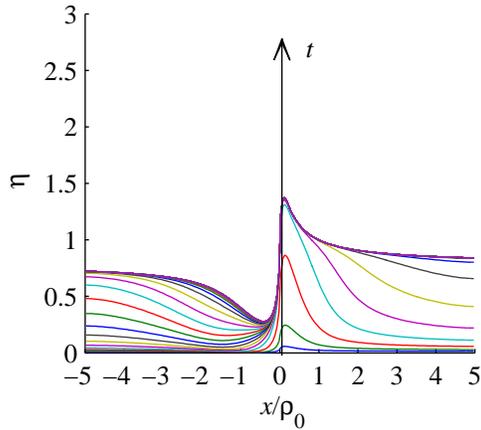}
\label{fig:ProfileY0Rneg1Upos1}
}
\subfigure[\hspace{1mm} $ r_0=-1$ and $u_1=-1$]{
\includegraphics[width=7cm]{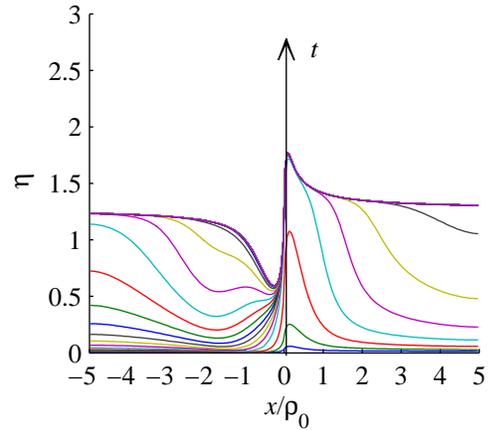}
\label{fig:ProfileY0Rneg1Uneg1}
}
\caption{Evolution of the order parameter $\eta(x/\rho_0,y=0)$  for different combinations of $r_{0}$ and $u_{1}$ with $v_{0}=1$. The time advances vertically in the plots; in a) to c) $\eta $ is given from $t=200\Delta t_{ref}\to 10000\Delta t_{ref}$, and in d) and e) from $t=50\Delta t_{ref}\to 2000\Delta t_{ref}$.}
\end{figure}

\begin{figure}[htbp]
\subfigure[\hspace{1mm} $ t=1000\Delta t _{ref} $]{
\label{fig:Surf-Time1000}
\includegraphics{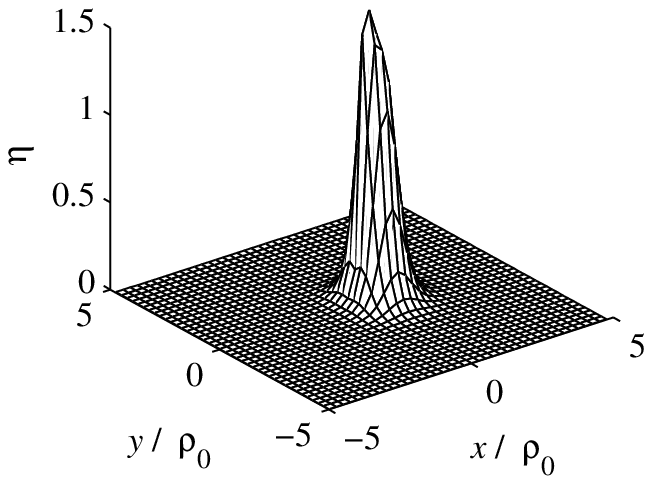}
}
\subfigure[\hspace{1mm} $ t = 2000\Delta t _{ref} $]{
\label{fig:Surf-Time2000}
\includegraphics{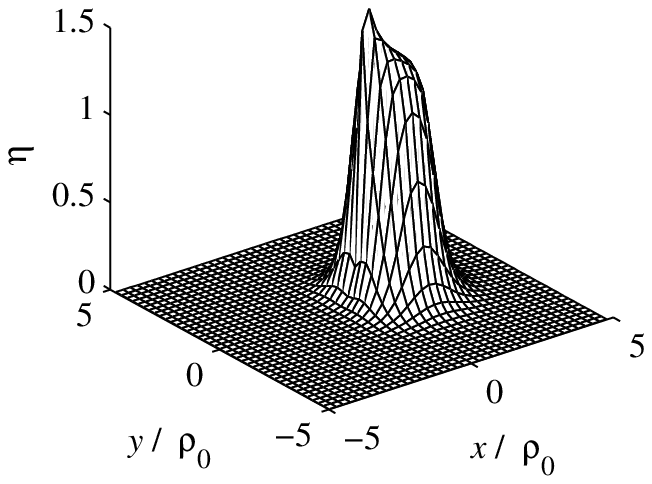}
}
\subfigure[\hspace{1mm} $ t=10000\Delta t _{ref} $]{
\label{fig:Surf-Time10000}
\includegraphics{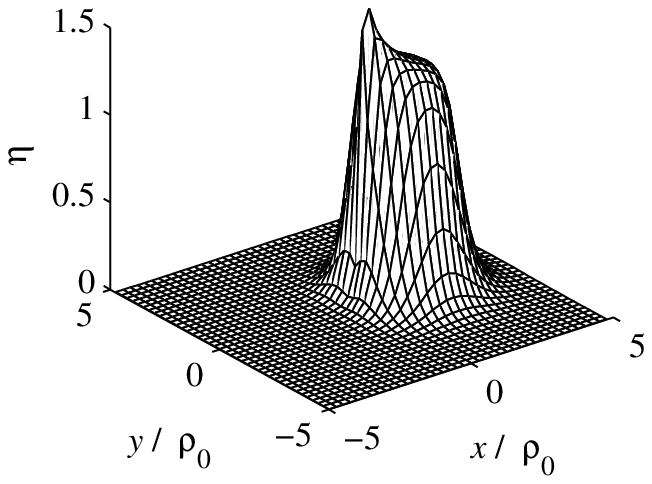}
}
\caption{Spatial distribution of $\eta$ at various times $t=$ 1000, 2000 and 10000~$\Delta t_{ref}$, for $r_{0}=1$, $u_{1}=-2$, and $v_{0}$. Only every fifth node in the mesh is selected for the illustration.}
\label{fig:SurfsRpos1Uneg2}
\end{figure}

\begin{figure}[htbp]
\subfigure[\hspace{1mm} $ t=1000\Delta t _{ref} $]{
\label{fig:SurfRpos1Uneg3Time1000}
\includegraphics{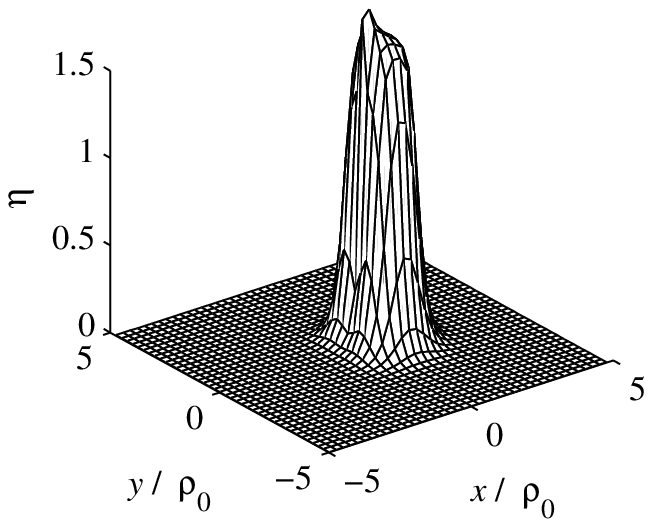}
}
\subfigure[\hspace{1mm} $ t = 2000\Delta t _{ref} $]{
\label{fig:SurfRpos1Uneg3Time2000}
\includegraphics{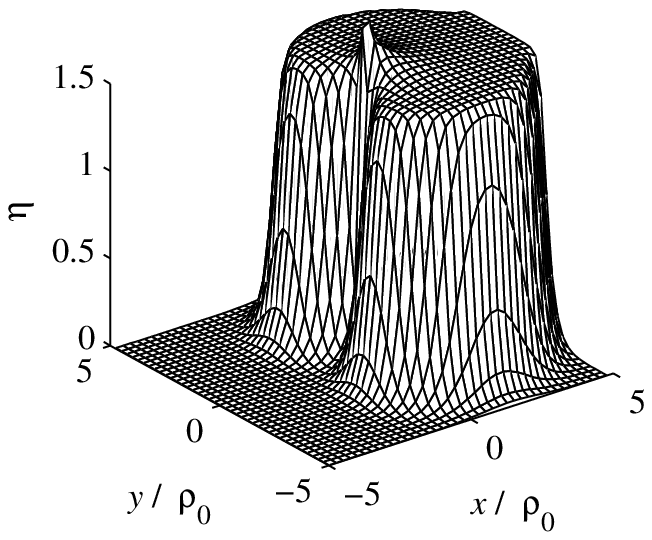}
} \\
 \subfigure[\hspace{1mm} $ t=6000\Delta t _{ref} $]{
 \label{fig:SurfRpos1Uneg3Time6000}
\includegraphics{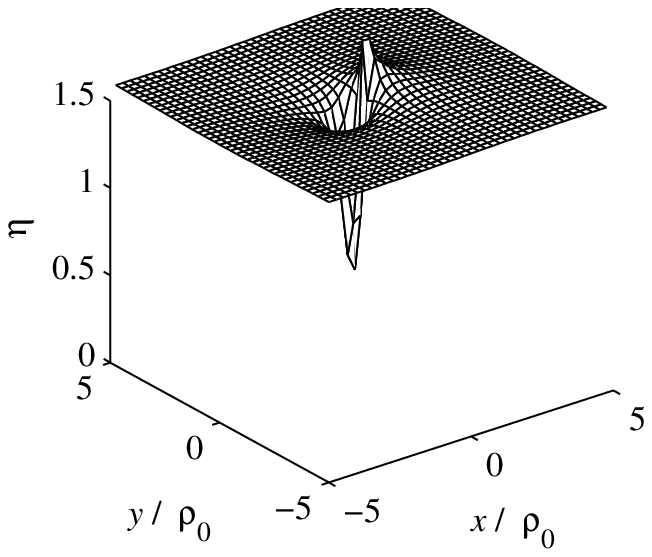}
}
\subfigure[\hspace{1mm} Contours at $t=6000\Delta t _{ref} $]{
\label{fig:ContRpos1Uneg3Time6000}
\includegraphics{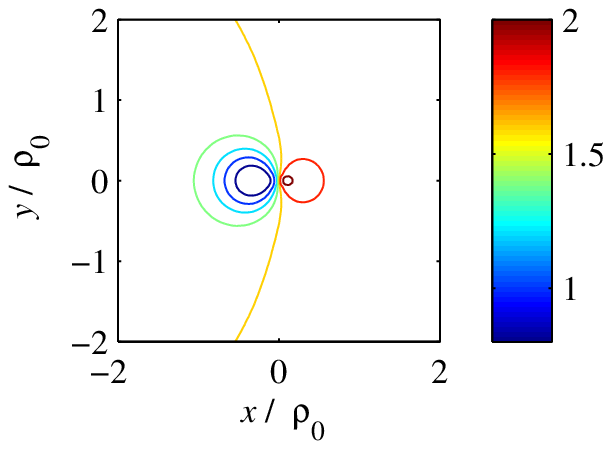}
}
\caption{Spatial distribution of $\eta$ at various times $t=$ 1000, 2000 and 6000~$\Delta t_{ref}$, for $r_{0}=1$, $u_{1}=-3$ and $v_{0}$. Only every fifth node in the mesh is seleted for the illustration.}
\label{fig:SurfsRpos1Uneg3}
\end{figure}

Defects shift the nucleation temperature locally to a higher value and make it space-dependent; see Eq. (\ref{eqn:tc-disl-2}). Let us first consider the situation with $T>T_{c0}$, i.e. $r_{0}>0$, and $u_{1}<0$. Figure~\ref{fig:SurfsRpos1Uneg1} shows the evolution of $\eta$ when $r_{0}=1$ and $u=-1$. As can be seen, a peak emerges in the $\eta$-surface, which is mainly situated on the compressive side of the dislocation ($x>0$), and it evolves until it finds a stable shape, i.e. a steady-state solution is obtained. In Fig.~\ref{fig:ProfileY0Rpos1Uneg1}, the temporal evolution of $\eta$ at $y=0$, i.e. for a cross-section along the $x$-axis, is shown at different times. A relatively sharp peak emerges first, and afterward some broadening occurs. The maximum value of this peak, $\eta_{max}=\mathrm{max}\,[\eta_{peak}]$, is reached at $t\approx1200\Delta t_{ref}$, and the steady-state shape comes after at $t\approx1800\Delta t_{ref}$. It is observed that after a relatively slow start, the evolution rate increases until the steady-state value, $\eta_{max}$, is nearly achieved. Afterward, it decreases just before the system reaches steady state. Next, the case with $u_{1}=-2$, $r_{0}=1$ is investigated, and the evolutions of the order parameter are presented in Figs.~\ref{fig:SurfsRpos1Uneg2} and \ref{fig:ProfileY0Rpos1Uneg2}. It is seen that a peak first develops reaching its maximum value, $\eta_{max}$, at $t\approx1200\Delta t_{ref}$, as in the case with $u_{1}=-1$. Then a considerable broadening takes place until a steady-state shape is reached at $t\approx7000\Delta t_{ref}$. We have also studied the situation where $u_{1}=-3$ and $r_{0}=1$. The $\eta$-surface at three different times displayed in Fig.~\ref{fig:SurfsRpos1Uneg3} and Fig.~\ref{fig:ProfileY0Rpos1Uneg3} shows $\eta(x,y=0)$ for
every 200 time-step. The order parameter evolution here differs from the cases with $u_{1}=-1$ and $u_1=-2$. Again a peak develops that afterward broadens, but now the second phase grows until the whole material is transformed to this phase except near the dislocation on its tensile side ($x<0$) where the transformation is held back. The peak reaches its maximum value, $\eta_{max}$, at $t\approx1200\Delta t_{ref}$, and steady state is obtained at $t\approx5000\Delta t_{ref}$. A close-up of the $\eta$ field, in the form of a contour plot, is shown in Fig.~\ref{fig:ContRpos1Uneg3Time6000}.

\begin{figure}[htbp]
\subfigure[\hspace{1mm} $ t=200\Delta t_{ref} $]{
\label{fig:SurfRneg1Upos1Time200}
\includegraphics{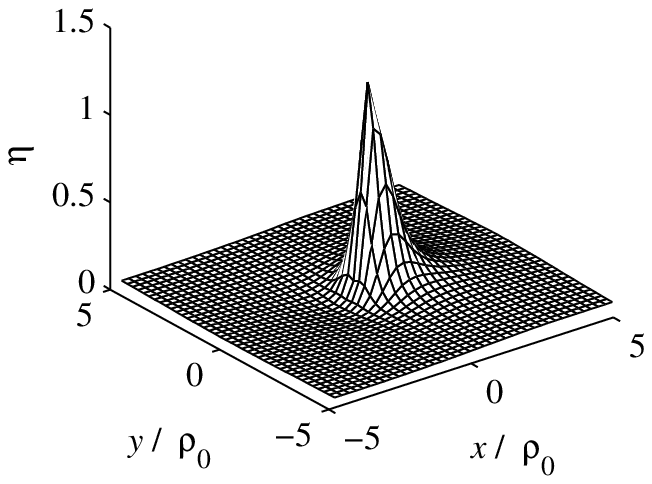}
}
\subfigure[\hspace{1mm} $ t=400\Delta t_{ref} $]{
\label{fig:SurfRneg1Upos1Time400}
\includegraphics{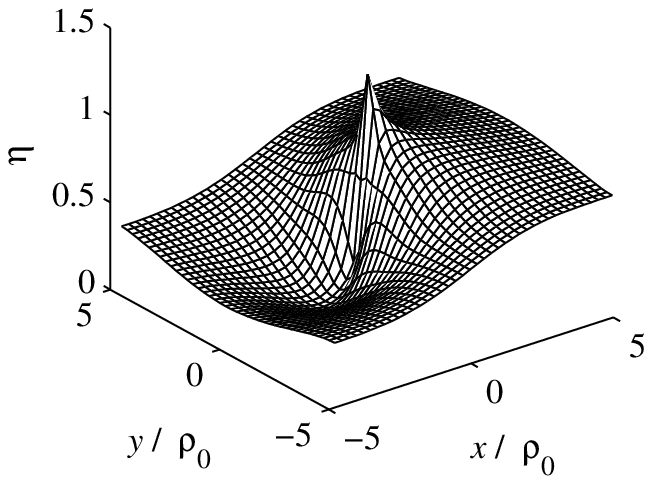}
}
\subfigure[\hspace{1mm} $ t = 2000\Delta t_{ref} $]{
\label{fig:SurfRneg1Upos1Time2000}
\includegraphics{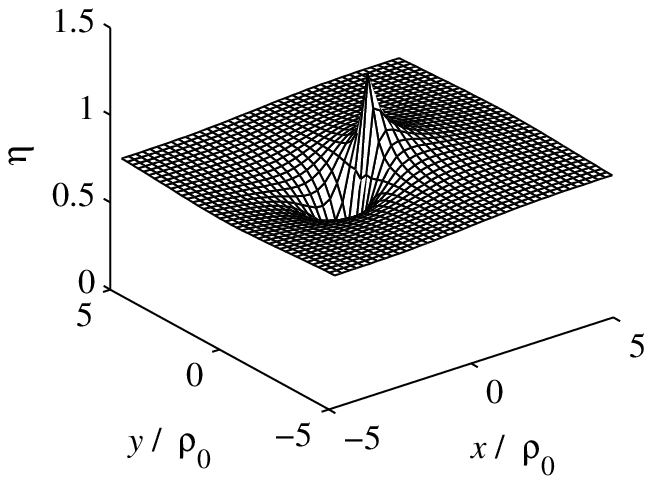}
}
\subfigure[\hspace{1mm} Contours at $ t = 2000\Delta t_{ref} $]{
\label{fig:ContRneg1Upos1Time2000}
\includegraphics{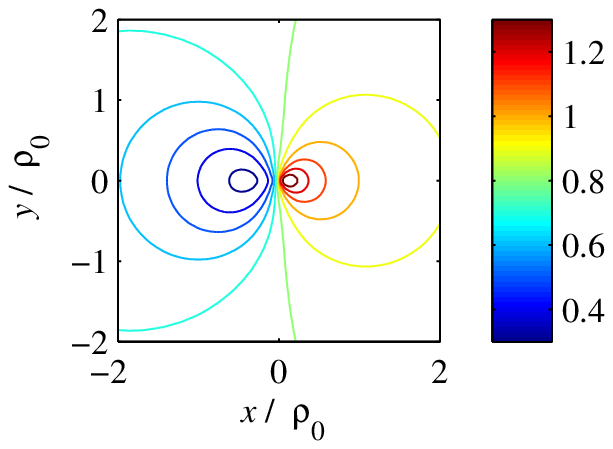}
}
\caption{Spatial distribution of $\eta$ at various times $t=$200, 400 and 2000 $\Delta t_{ref}$, for $r_{0}=-1$, $u_{1}=1$ and $v_{0}$. Only every fifth node in the mesh is selected for the illustration.}
\label{fig:SurfRneg1Upos1Time}
\end{figure}

The case $r_{0}=-1$ $(T<T_{c0})$ corresponds to a quick decrease of temperature, i.e. a quench below the transition temperature of the undisturbed system. At the same time as the temperature changes, an edge dislocation is introduced. First, the results from computations with $u_{1}=1$ are presented. As can be seen clearly from  Fig.~\ref{fig:SurfRneg1Upos1Time}, a peak is growing where compressive stresses are induced by the dislocation (i.e. $x>0$) with its maximum close to the dislocation. The evolution of $\eta$ is also shown in Fig. \ref{fig:ProfileY0Rneg1Upos1} as profiles of $\eta(x,y=0)$. Here, the peak evolves until $t\approx300\Delta t_{ref}$. Subsequently, $\eta$ increases in the entire material, while leaving a valley on the tensile side of the dislocation. The time to reach steady state is found to be \emph{$t\approx900\Delta t_{ref}$}.

For $r_{0}=-1$ and $u_{1}=-1$, the development of $\eta$ has the same character as with $u_{1}=1$, cf. Fig.~\ref{fig:ProfileY0Rneg1Uneg1}. The steady-state value of $\eta$ is, however, larger in the whole plane, while maximum values are reached at approximately the same time as with $u_{1}=1$. Steady state seems to be achieved at $t\approx900\Delta t_{ref}$. The steady-state solutions for $r_{0}=-1$ is similar to that of $r_{0}=1$ and $u_{1}=-3$. However, nucleation of second phase would not initiate in the tensile region of the solid. Instead, for $r_{0}<0,$ after the local phase transition near the dislocation, the order parameter $\eta$ increases in the whole material at the same time.

To recap, the values of some characteristic measures for the studied cases are given in Table \ref{table:CompRes}. It is seen that $\eta_{max}$ increases with decreasing $u_{1}$ for both positive and negative values of $r_{0}$. Equation \ref{eqn:u1-elastic} tells us that $u_{1}$ decreases by increasing the interaction strength $\alpha$, i.e. the interaction between $\eta$ and $\mathbf{u}$. Thus, with a larger value of $\alpha$, the singularity of the stress field induced by the dislocation, contributes to a larger value of $\eta$. Also, decreasing the elastic moduli, i.e. decreasing $\Lambda$, reduces $u_{1}$. The time needed for $\eta_{peak}$ to reach its maximum  $\eta_{max}$ is denoted by $t_{mp}$. In the case of $r_{0}=1$, a top that emerges near the dislocation is found to grow four times slower than in the case of quenching, i.e. with $r_{0}=-1$. This could be explained by indicating that in the quenching situation a defect-free material is expected to fully transform into the second phase, and thus the driving force for phase transition is greater for a system with $r_{0}<0$. Correspondingly, the time to reach steady state, $t_{ss}$, is shorter with $r_{0}=-1$ than with $r_{0}=1$ for $u_1=-1$. For the latter case, the evolution of a relatively thin top is faster than for lower values of $u_{1}$. With $u_{1}=-2$, a larger area at the dislocation transforms and it takes about four times longer. When practically the whole material is transformed into the second phase, $u_{1}=-3$, the growth rate increases, which may be expected. To estimate $\eta$ as $\rho\rightarrow\infty$, the value at $(x,y)=(5\rho_{0},5\rho_{0})$ is used as a measure and is denoted by $\eta_{\infty}$. The results are discussed further in subsection \ref{sec:SSResults}.

%
\begin{table}[b]
\caption{\label{table:CompRes}
Comparison between  results \{$\eta_{max}$, $t_{mp}$, $t_{ss}$, $\eta_{\infty}$\}
for different input values \{$r_{0}$, $u_{1}$\}.}
\begin{ruledtabular}
\begin{tabular}{cccccc}
 $r_0 \Rightarrow$  & $1$  &  $1$  & $1$  & $-1$ & $-1$ \\
 $u_1 \Rightarrow$  & $-1$ &  $-2$ & $-3$ & $1$  & $-1$ \\
\hline
$\eta_{max}$ & 1.46 & 1.78 & 2.05 & 1.37 & 1.77\\
$t_{mp}/\Delta t_{ref}$ & 1200 & 1200 & 1200 & 300 & 300\\
$t_{ss}/\Delta t_{ref}$ & 1800 & 7000 & 5000 & 900 & 900\\
$\eta_{\infty}$ & 0 & 0 & 1.63 & 0.81 & 1.29\\
\end{tabular}
\end{ruledtabular}
\end{table}



\subsection{Nucleus shape evolution}
\label{sec:NuclShape}

\begin{figure}[htbp]
\subfigure[\hspace{1mm} $t=1000\Delta t_{ref} $]{
\label{fig:BWRpos1Uneg2Time1000}
\includegraphics[width=4cm]{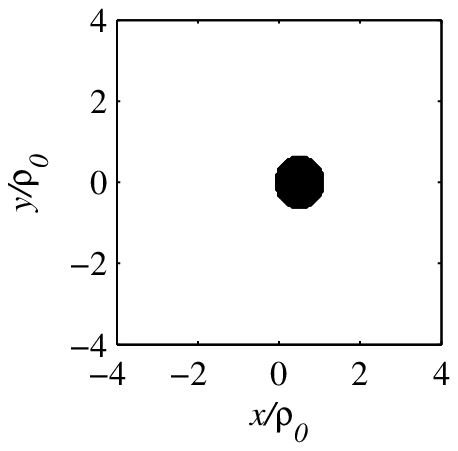}
}
\subfigure[\hspace{1mm} $ t=2000\Delta t_{ref} $]{
\label{fig:BWRpos1Uneg2Time2000}
\includegraphics[width=4cm]{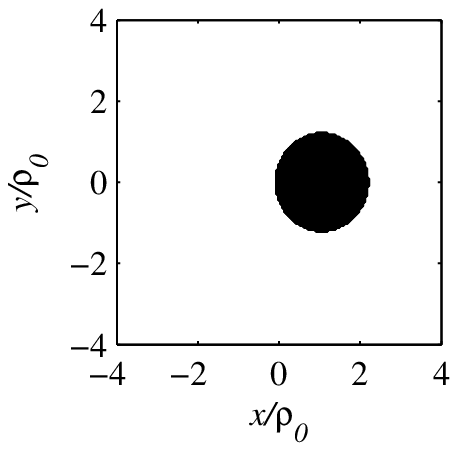}
}
\subfigure[\hspace{1mm} $ t = 10000\Delta t_{ref} $]{
\label{fig:BWRpos1Uneg2Time10000}
\includegraphics[width=4cm]{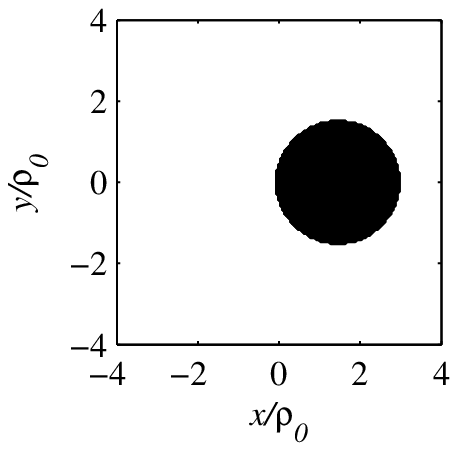}
}
\subfigure[\hspace{1mm} $ t=1000\Delta t_{ref} $]{\label{fig:BWRpos1Uneg3Time1000}
\includegraphics[width=4cm]{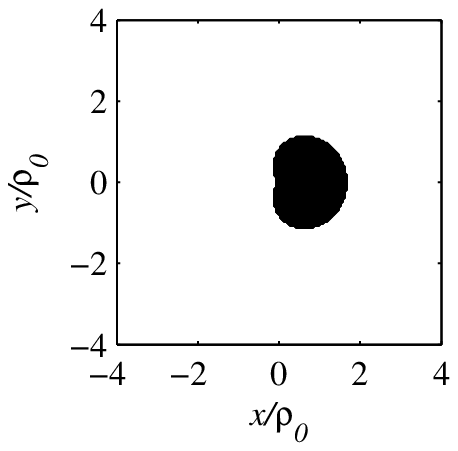}
}
\subfigure[\hspace{1mm} $ t=2000\Delta t_{ref} $]{
\label{fig:BWRpos1Uneg3Time2000}
\includegraphics[width=4cm]{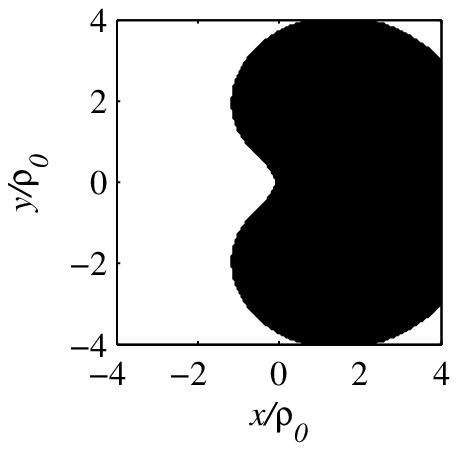}
}
\subfigure[\hspace{1mm} $ t = 6000\Delta t_{ref} $]{
\label{fig:BWRpos1Uneg3Time6000}
\includegraphics[width=4cm]{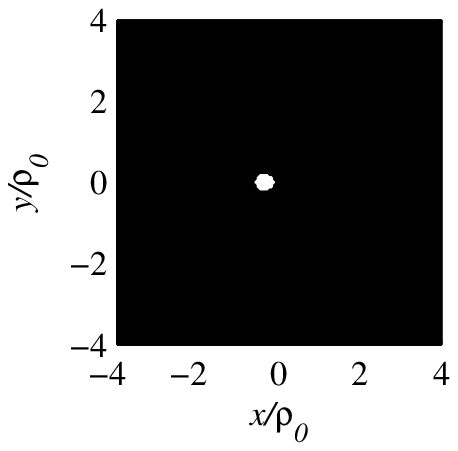}
}
\subfigure[\hspace{1mm} $ t=400\Delta t_{ref} $]{
\label{fig:BWRneg1Upos1Time400}
\includegraphics[width=4cm]{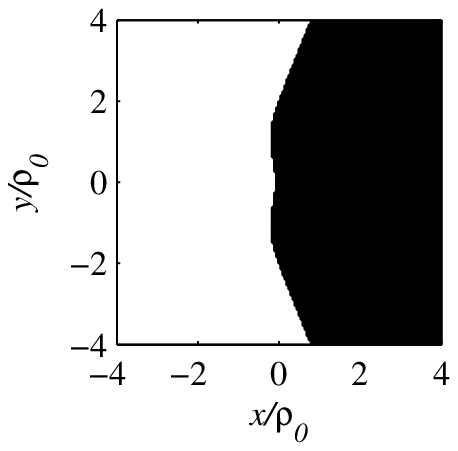}
}
\subfigure[\hspace{1mm} $ t=550\Delta t_{ref} $]{
\label{fig:BWRneg1Upos1Time550}
\includegraphics[width=4cm]{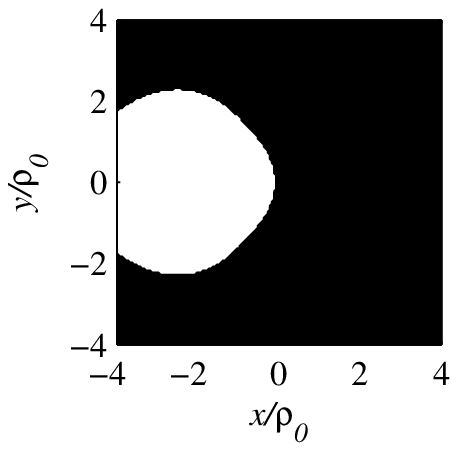}
}
 \subfigure[\hspace{1mm} $ t = 2000\Delta t_{ref} $]{
\label{fig:BWRneg1Upos1Time2000}
\includegraphics[width=4cm]{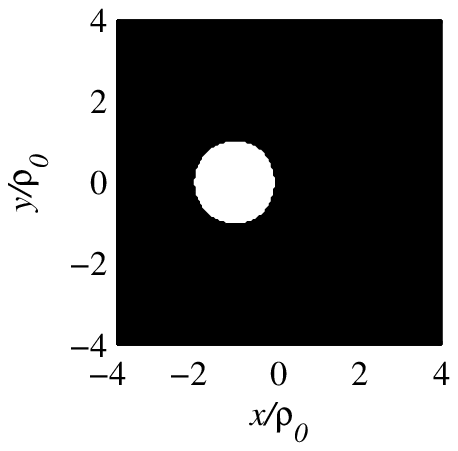}
} \\
\subfigure[\hspace{1mm} $ t=400\Delta t_{ref} $]{
\label{fig:BWRneg1Uneg1Time400}
\includegraphics[width=4cm]{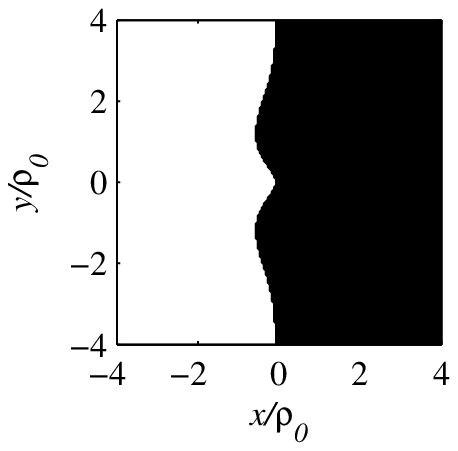}
}
\subfigure[\hspace{1mm} $ t=550\Delta t_{ref} $]{
\label{fig:BWRneg1Uneg1Time550}
\includegraphics[width=4cm]{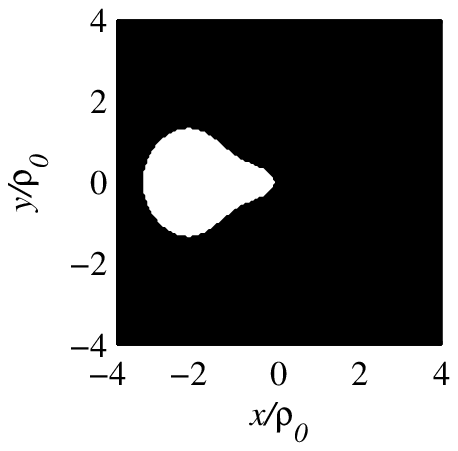}
}
\subfigure[\hspace{1mm} $ t = 2000\Delta t_{ref} $]{
\label{fig:BWRneg1Uneg1Time2000}
\includegraphics[width=4cm]{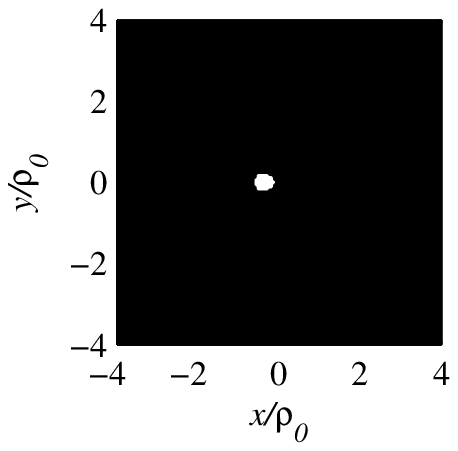}
}
\caption{Snapshots of evolution of a second phase (black) from the matrix material (white) for the cases (a)-(c) $r_{0}=1$, $u_{1}=-2$, (d)-(f) $r_{0}=1$, $u_{1}=-3$, (g)-(i) $r_{0}=-1$, $u_{1}=1$, and (j)-(l) $r_{0}=-1$, $u_{1}=-1$. }
\label{fig:BWWithDisl}
\end{figure}

To illustrate the evolution of a second phase in the $(x,y)$-plane, black and white patterns representing the two phases are displayed in Fig. \ref{fig:BWWithDisl}. Black indicates areas where $\eta>\eta_{max}/2$ and represents the second phase and vice verse. This choice is somewhat arbitrary, but renders patterns that illustrate the evolution of the shape of a nucleus. Patterns for four different combinations of $r_{0}$ and $u_{1}$ at three different times are given in the figure. In the upper row, Figs. \ref{fig:BWRpos1Uneg2Time1000} to \ref{fig:BWRpos1Uneg2Time10000}, patterns for the case $r_{0}=1$ and $u_{1}=-2$ are shown. It is clearly seen that the second phase evolves in the shape of a circle and is located on the compressive side of the dislocation, i.e. where $x > 0$. For $u_{1}=-3$, the growing second phase is not circular as in the other cases. Instead, it is kidney shaped, see Fig.~\ref{fig:BWRpos1Uneg3Time1000} and \ref{fig:BWRpos1Uneg3Time2000} in the second row. As steady state is reached, see Fig.~\ref{fig:BWRpos1Uneg3Time2000}, the area where the transformation is held back by the dislocation is circular.

When $r_{0}=-1$, the patterns for $u_{1}=1$ and $u_{1}=-1$ are shown in the two lower rows of Fig. \ref{fig:BWWithDisl}, respectively. It is seen that the second-phase evolution is somewhat slower for $x<0$. The only significant difference in the behavior is the way that the matrix phase shrinks; cf. Figs.~\ref{fig:BWRneg1Upos1Time400} to \ref{fig:BWRneg1Upos1Time2000} with \ref{fig:BWRneg1Uneg1Time400} to \ref{fig:BWRneg1Uneg1Time2000}. In steady state, the area where the transformation is held back is circular in both cases, although it is smaller with the negative $u_{1}$.



\subsection{Steady-state solutions}
\label{sec:SSResults}

Let us consider the long-time limit when steady state is reached, i.e. $\partial\eta/\partial t=0$. Furthermore, we assume the system is away from the critical region, where the long wavelength fluctuation in the order parameter can be neglected, i.e., $g_{a}\nabla^{2}\eta\approx0$. Then ignoring the noise term, Eq.~(\ref{eqn:tdgl-eq2}) gives:
\begin{equation}
\bar{\eta}\left(r_{1}+u_{1}\bar{\eta}^{2}+v_{0}\bar{\eta}^{4}\right)=0,
\label{eq:redGovEq}
\end{equation}
where $\bar{\eta}$ is the elastic medium (mean-field) order parameter containing the edge dislocation. This reduced equation corresponds to $d\mathcal{V}(\bar{\eta})/d\bar{\eta}=0$ with modified coefficients $r_{1}$ and $u_{1}$ instead of $r_{0}$ and $u_{0}$, cf. Eqs.~(\ref{eqn:toten1}) to (\ref{eqn:u1-elastic}). The non-zero solutions of Eq. (\ref{eq:redGovEq}) are
\begin{equation}
\bar{\eta}_{\pm}^{2}=\frac{-u_{1}\pm\sqrt{u_{1}^{2}-4v_{0}r_{1}}}{2v_{0}}.
\label{eq:etaSq}
\end{equation}
The solutions $\bar{\eta}_{+}$ exist for $r_{1}\le u_{1}^{2}/4v_{0}$, whereas those of $\bar{\eta}_{-}$ exist for $0<r_{1}\le u_{1}^{2}/4v_{0}$. We also note that $\mathcal{V}(\bar{\eta})=0$ together with d$\mathcal{V}(\bar{\eta})/d\bar{\eta}=0$ yield $r_{1}=3u_{1}^{2}/16v_{0}$.

\begin{figure}[htbp]
\subfigure[\hspace{1mm} $u_1 > 0$]{
\label{fig:LandauUp}
\includegraphics{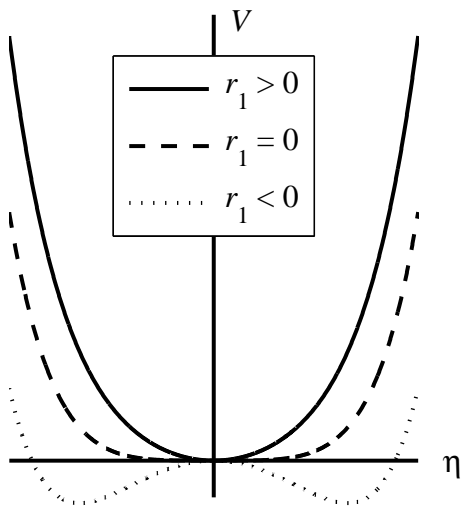}
}
\hspace{1cm}
 \subfigure[\hspace{1mm} $u_1 < 0$]{
 \label{fig:LandauUn}
\includegraphics{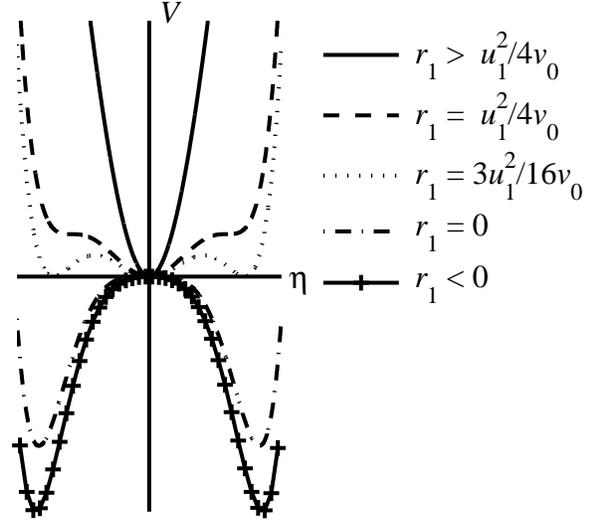}
}
\caption{The Landau potential $\mathcal{V}(\eta)=\frac{1}{2}r_{1}\eta^{2}+\frac{1}{4} u_{1}\eta^{4}+\frac{1}{6}v_{0}\eta^{6}$ with $v_{0}>0$ versus $\eta$ for various combinations of the coefficients.}
\label{fig:Landau-pot}
\end{figure}

Figure~\ref{fig:Landau-pot} illustrates two different settings of the the coefficients of $\mathcal{V}(\eta)$: (a) $u_{1}>0$, and (b) $u_{1}<0$, both with $v_{0}>0$. If $u_{1}>0$ and $r_{1}>0$, no nucleation of second phase would occur, whereas $r_{1}=0$ sets off the nucleation and $r_{1}<0$ corresponds to formation and an unstable growth state of the second phase. In the case of $u_{1}<0$, at high temperatures, $r_{1}$ is large and positive and $\mathcal{V}(\bar{\eta})$ has a simple structure with a minimum at $\bar{\eta}=0$. As the temperature is decreased $r_{1}$ gets smaller and in the range $r_{1}<u_{1}^{2}/4v_{0}$, $u_{1}<0$ spawns two local minima at $\bar{\eta}\ne0$, i.e. emergence of metastable second-phase states, while the solid solution is stable (the global minimum). A further decrease of temperature makes $r_{1}<3u_{1}^{2}/16v_{0}$, at which the global minimum shifts from $\bar{\eta}=0$ to the two symmetrically located states with $\bar{\eta}\ne0$. This marks a first-order phase transition. For $r_{1}<0$ the local stability of the solution $\bar{\eta}=0$ disappears.

\begin{figure}[htbp]
\subfigure[\hspace{1mm} $ r_0=1, u_1=1$]{
\label{fig:CompSSR1U1}
\includegraphics{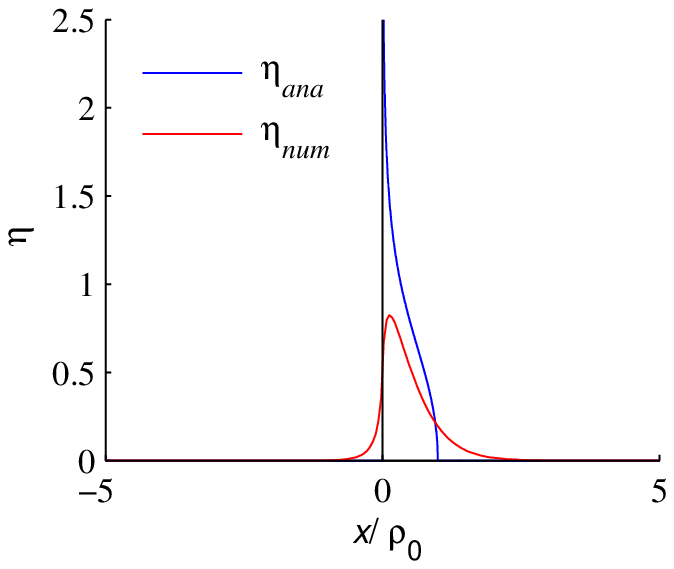}
}
\subfigure[\hspace{1mm} $ r_0=1, u_1=-1$]{
\label{fig:CompSSR1U-1}
\includegraphics{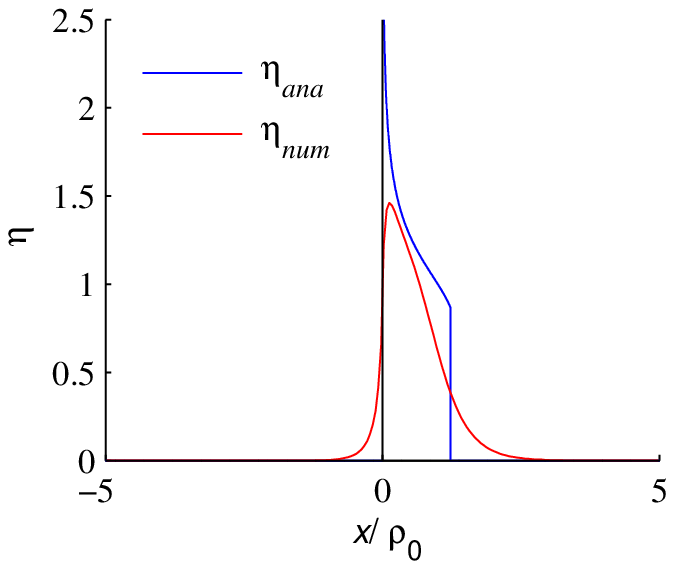}
}
\subfigure[\hspace{1mm} $ r_0=1, u_1=-2$]{
\label{fig:CompSSR1U-2}
\includegraphics{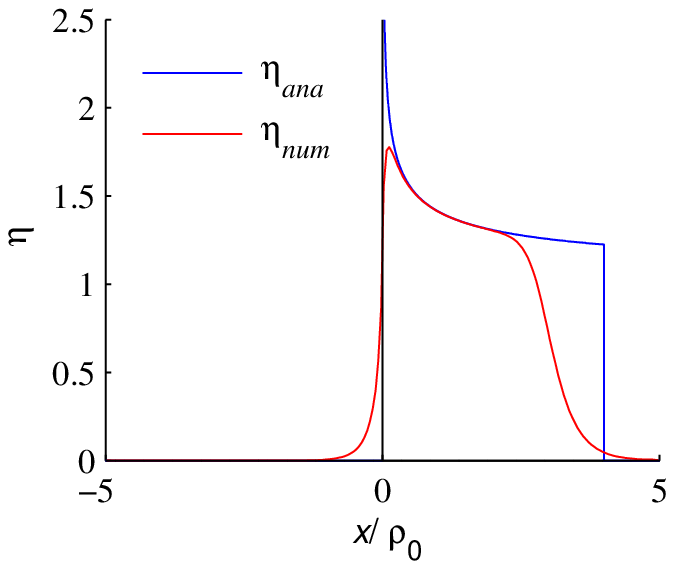}
}
\subfigure[\hspace{1mm} $ r_0=1, u_1=-3$]{
\label{fig:CompSSR1U-3}
\includegraphics{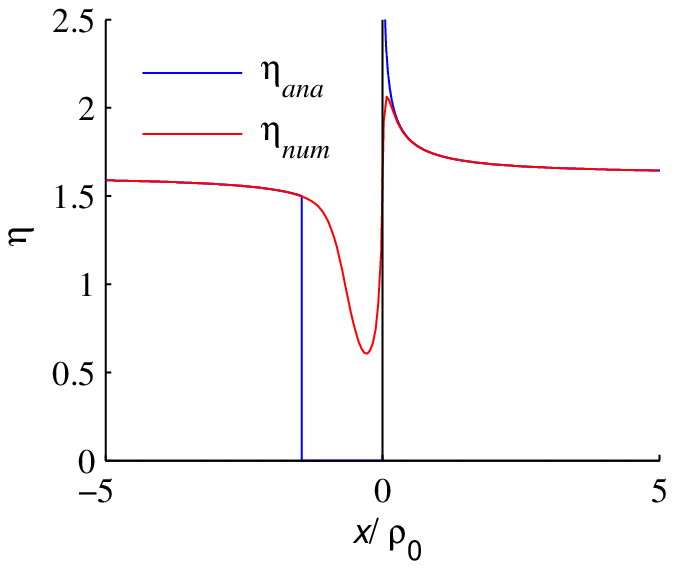}
}
\subfigure[\hspace{1mm} $ r_0=-1, u_1=1$]{
\label{fig:CompSSR-1U1}
\includegraphics{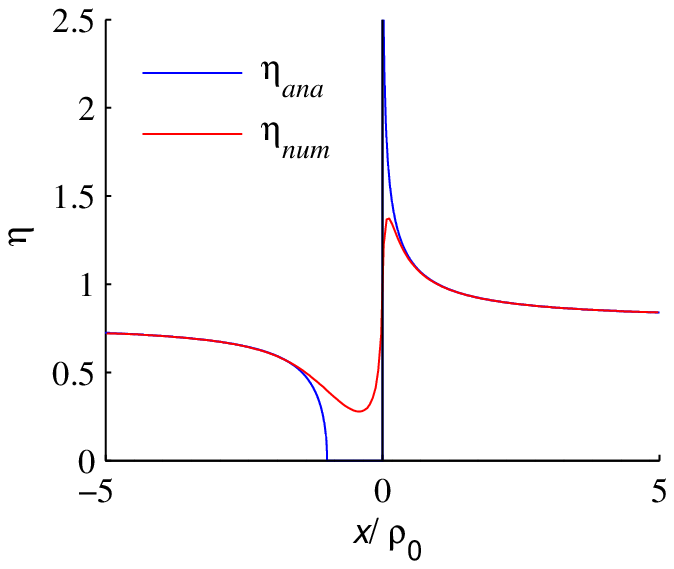}
}
\subfigure[\hspace{1mm} $ r_0=-1, u_1=-1$]{
\label{fig:CompSSR-1U-1}
\includegraphics{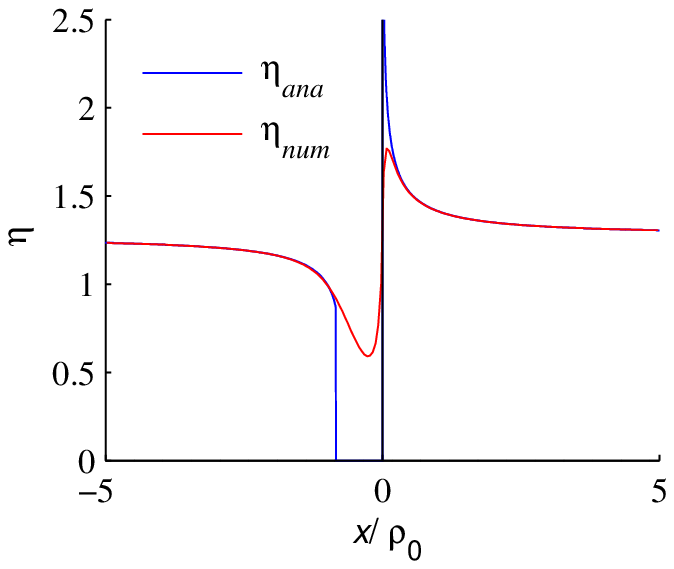}
}
\caption{Comparison between analytical and numerical computations for spatial variation $(x,y=0)$ of order parameter in steady state
for different parameter sets ($r_{0}$, $u_{1}$) with $v_{0}=1$.}
\label{fig:CompSS}
\end{figure}

The parameter $r_{1}$ is a function of both temperature and spatial coordinates. Accordingly, so are $\bar{\eta}_{+}$ and the location where nucleation and further evolution of the second phase would occur. Depending on the sign of $u_{1}$, different situations would arise. Figure~\ref{fig:CompSS} shows $\eta$ along the $x$-axis obtained by the numerical computations, $\eta_{num}$, which includes the term  $g_{a}\nabla^{2}\eta$ with  $g_{a}=0.1$, and the aforementioned analytical solution, $\eta_{ana}\equiv\bar{\eta}$, for different parameter settings. We see that the main features of $\eta_{ana}$ can also be found from the numerical solutions. The deviations are largest near the dislocation line, where $\eta_{ana}$ goes to infinity, and at the locations where there are sharp boundaries between phases. This is expected since the phase-field model (numerical solution) produces smooth interfaces. For situations with $u_{1}>0$, regardless of magnitude, Eq.~(\ref{eqn:r1-disl}) tells us that the ordered phase would evolve within a circle confined by $\rho=\rho_0\cos{\theta}$, if $r_{0}>0$. In the quenched case ($r_{0}<0)$, the whole material would transform except where $\rho<-\rho_0\cos{\theta}$, i.e. inside a circle of a radius $\rho_0$. The values of $u_{1}$ would only influence the magnitude of $\bar{\eta}_{+}$. The profiles $\eta_{ana}$ in Figs.~\ref{fig:CompSSR1U1} and \ref{fig:CompSSR-1U1} show that boundaries between phases in both cases are continuous (second-order transition). For negative $u_{1}$, the radii of the limiting circles vary with $u_{1}$  as $\rho_0/(1-3u_1^2/16)$, since a first-order transition occurs if $r_{1}<3u_{1}^{2}/16v_{0}$. In Figs.~\ref{fig:CompSSR1U-1} to \ref{fig:CompSSR1U-3} and \ref{fig:CompSSR-1U-1}, it is seen that the order parameter is discontinuous at the corresponding locations.

We should note that the smoothness of the interfaces in the numerical computations is governed by the term $g_{a}\nabla^{2}\eta$ in Eq.~(\ref{eqn:tdgl-eq2}). For instance, if the value of the coefficient $g_{a}$ is reduced by two orders of magnitude, the softness of the phase interface is reduced substantially, giving rise to a sharper shape that is closer to $\eta_{ana}$.


\subsection{Influence of periodic boundary conditions}
\label{sec:InflPerBC}

In the case of $r_{0}=1$ and $u_1=-3$, cf. Fig.~\ref{fig:SurfRpos1Uneg3Time6000} and \ref{fig:CompSSR1U-3}, the periodic boundary conditions inevitably have an impact on the second-phase evolution; the broadening of the top continues until the boundary is reached at $t\approx1800\Delta t_{ref}$. This means that there is an interaction with the corresponding dislocations that are mirrored at the boundaries, i.e. those located at $(x,y)=\pm10\rho_{0}$. However, if the mesh size is increased, thus increasing the distance to the image dislocations, the same behavior is obtained. That is, no significant difference in growth rates is found, except that the time that would take for the second phase to come into contact with the edge of the grid would be twice as long. The relatively large difference between $\eta_{ana}$ and $\eta_{num}$ for the case $r_0=1$ and $u_{1}=-2$ is probably due to  the imposed periodic boundary conditions, but that impact was not investigated further.


\section{Influence of Thermal noise}
\label{sec:Noise}

In the preceding computations, the Langevin thermal noise, i.e. the $\vartheta_{a}$ term on the right hand side in Eq.~(\ref{eqn:tdgl-eq2}), was disregarded. In order to investigate the influence of this term on the temporal evolution of the system, numerical computations have been performed for situations with and without the presence of a dislocation. The thermal noise is represented by a Gaussian distribution with the mean value equal to zero and the variance of $10^{-4}$. Instead of using a uniform random distribution as the initial value for $\eta$, cf. in section~\ref{sec:Numeric}, a constant initial value equal to $7.5\times10^{-3}$ is used. Then for each time increment a new noise distribution is added. Here, we have used a finer mesh, namely a mesh with $400\times400$ elements, compared to our foregoing computations, and also the time increment is taken to be smaller than before, i.e., only a fourth of the previous one: $\Delta t=\Delta t_{ref}/4$.


\subsection{System without a dislocation}
\label{sec:NoiseNoDisl}

\begin{figure}[htbp]
\subfigure[\hspace{1mm} $25 \Delta t_{ref}$]{
\includegraphics{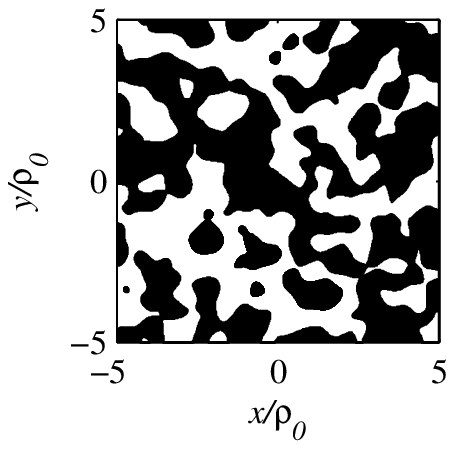}
\label{fig:PatternNoDislTime25}
}
\subfigure[\hspace{1mm} $50 \Delta t_{ref}$]{
\includegraphics{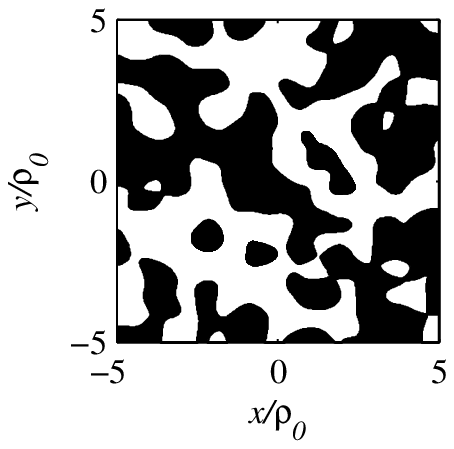}
\label{fig:PatternNoDislTime50}
}
\subfigure[\hspace{1mm} $200 \Delta t_{ref}$]{
\includegraphics{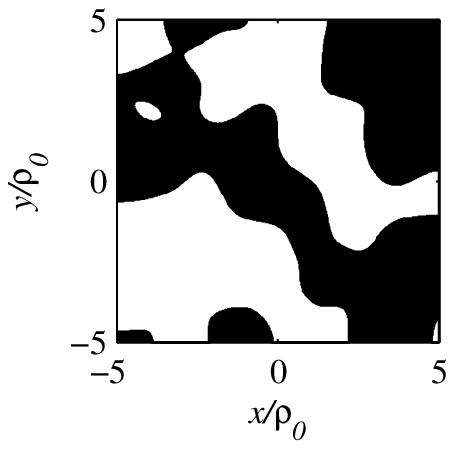}
\label{fig:PatternNoDislTime200}
}
\subfigure[\hspace{1mm} $25 \Delta t_{ref}$]{
\includegraphics{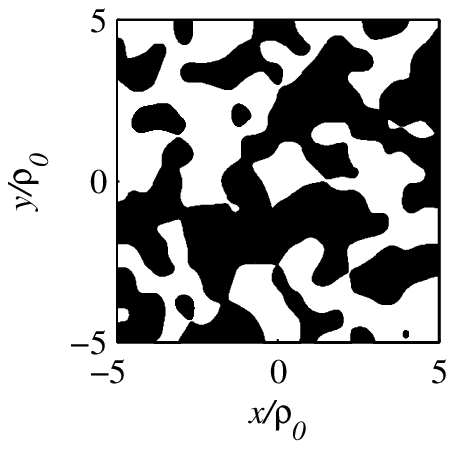}
\label{fig:PatternNoDislNoiseTime25}
}
\subfigure[\hspace{1mm} $50 \Delta t_{ref}$]{
\includegraphics{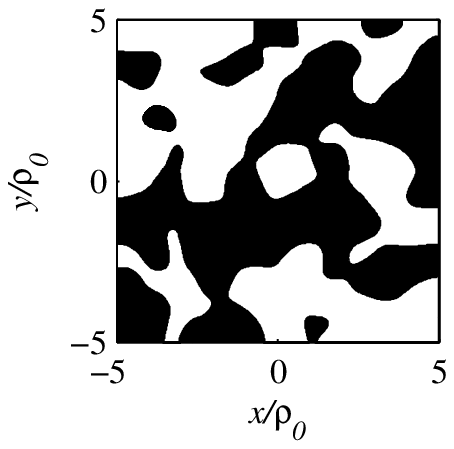}
\label{fig:PatternNoDislNoiseTime50}
}
\subfigure[\hspace{1mm} $200 \Delta t_{ref}$]{
\includegraphics{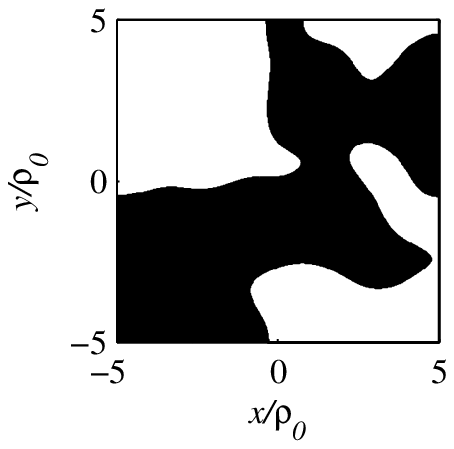}
\label{fig:PatternNoDislNoiseTime200}
}
\caption{Snapshots of domain coarsening on quenching for the case of  no dislocation. Top panel: without thermal noise term; bottom: with the thermal noise term. Here, $r_{0}=-1,u_{1}=-1,v_{0}=1$.}
\label{fig:PatternNoDislNoise}
\end{figure}

\begin{figure}[htbp]
\includegraphics{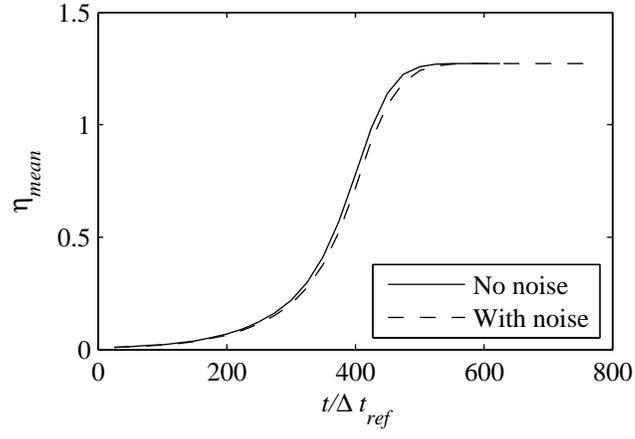}
\caption{Temporal evolution of the system without a dislocation: mean field order parameter $\eta_{mean}$ versus  scaled time $t/\Delta t_{ref}$ with and without the presence of thermal noise ($r_{0}=-1,u_{1}=-1,v_{0}=1$).}
\label{fig:CompNoiseMeanEtaVsTime}
\end{figure}

For reference, a system without a dislocation and $u_{1}=-1$ is studied. The system is quenched from above to below the critical temperature $T_c$ instantaneously, i.e.  from  $r_{0}>0$ to $r_{0}<0$. In such a case the whole system is expected to transform into the second phase, see e.g. Fig. \ref{fig:LandauUn}. First a computation without the thermal noise term is performed. The top row of Fig.~\ref{fig:PatternNoDislNoise} shows the evolving domain at three different periods. It is clearly seen that the domain is coarsening with time. We have not yet quantified the level of coarseness as time advances. It should be mentioned that the white and black areas do not represent the matrix phase and second phase, respectively, instead a mean value of $\eta$ over the whole area, $\eta_{mean}$, at each time step is determined, and locations with $\eta$ greater than this mean value are indicated in black, and viceversa. Since the GLK equation is non-conservative, the entire system is eventually engulfed by the second phase. From the plot of  $\eta_{mean}$ versus $t$ the variation in the rate of transformation can be inferred, Fig.~\ref{fig:CompNoiseMeanEtaVsTime}. It is seen that after an initial period with quite a slow increase in $\eta$, the transformation accelerates until the equilibrium value is reached. The curves in Fig. \ref{fig:CompNoiseMeanEtaVsTime} fit a logistic curve of the form
\begin{equation}
\eta(t) \approx \frac{a}{1+b\exp(-c t)}  ,
\label{eq:logisic-fit}
\end{equation}
with $a=1.289,b=5486.75$, and $c=0.02242$. The fit is quite good for $t/\Delta t_{ref} \ge 400$.

The corresponding calculation is done with the presence of thermal noise as described above. In Fig.~\ref{fig:PatternNoDislNoise}, the bottom row shows the obtained patterns for the same times as for the case without noise. A somewhat coarser domain is obtained for the situation with thermal noise. No significant difference in the rate of phase transformation is, however, found, see Fig.~\ref{fig:CompNoiseMeanEtaVsTime}. Based on these observations, we conclude that the influence of thermal noise is insignificant for the domain growth on quenching in the absence of dislocation, at least for the parameter set used in our comparative study. We shall recall, though, that the absence of the thermal noise term in Eq.~(\ref{eqn:tdgl-eq2}) assumes that the system is effectively evolving at $T=0$ K. In this context, it has been pointed out that the renormalization group (RG) theory shows three fixed points for temperature, namely,   $T=0$, $T=T_c$ and  $T=\infty$ \cite{Bray_1994}. This implies that during ordering, temperatures below $T_c$  move toward zero, while those  under disordering move toward infinity in RG flow maps. Therefore, the final quenching temperature $T_F$ is an impertinent variable for quenches into the ordered phase, which justifies the neglect of thermal noise in the equations of motion  \cite{Bray_1994}.

The domain shapes observed in Fig.~\ref{fig:PatternNoDislNoise} are equivalent to the snapshots of coarsening produced by models with non-conserved order parameter dynamics, e.g. the kinetic Ising or KIM model  \cite{Bray_1994,Olejarz_et_al_2012}, which is in the same university class as the GLK model employed here. In more detail, the recent publication by Olejarz and colleagues \cite{Olejarz_et_al_2012}, who used a 2D KIM model on a 1024$\times$1024 square lattice with periodic boundary conditions, show very similar snapshots of coarsening as our Fig.~\ref{fig:CompNoiseMeanEtaVsTime}, after a quench from $T=\infty$  to $T=0$. This again vindicates our conclusion regarding the effect of thermal noise on the kinetics of domain growth. The time evolution patterns in Fig. \ref{fig:PatternNoDislNoise} may also be compared with those produced in Onuki's book \cite{Onuki_2002}, where the results of an exact numerical solution of model A (with cubic order parameter) are compared with other known approximate solutions in the literature.


\subsection{System with a dislocation}
\label{sec:NoiseWithDisl}

\begin{figure}[htbp]
\subfigure[\hspace{1mm} $25\Delta t_{ref}$]{
\includegraphics{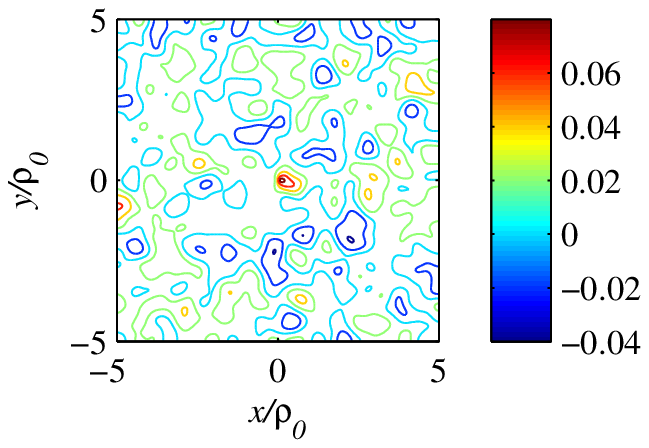}
\label{fig:PatternNoiseR1Time25}
}
\subfigure[\hspace{1mm} $50\Delta t_{ref}$]{
\includegraphics{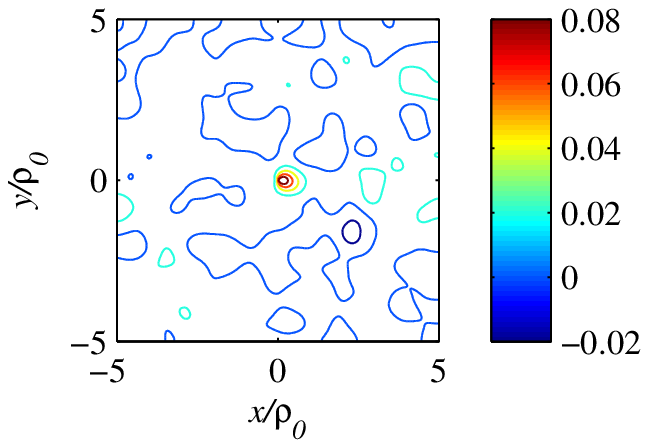}
\label{fig:PatternNoiseR1Time50}
}
\subfigure[\hspace{1mm} $200\Delta t_{ref}$]{
\includegraphics{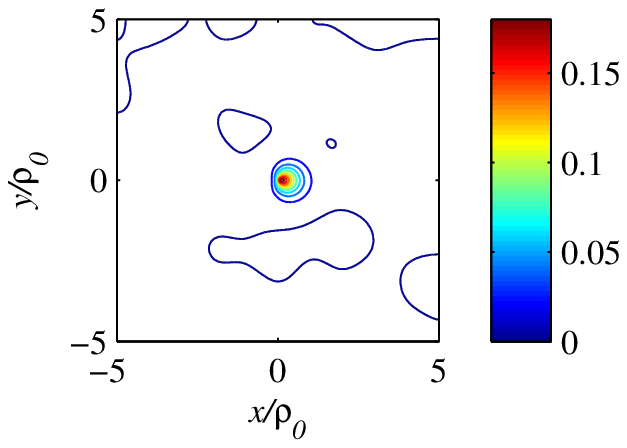}
\label{fig:PatternNoiseR1Time200}
}
\subfigure[\hspace{1mm} $500\Delta t_{ref}$]{
\includegraphics{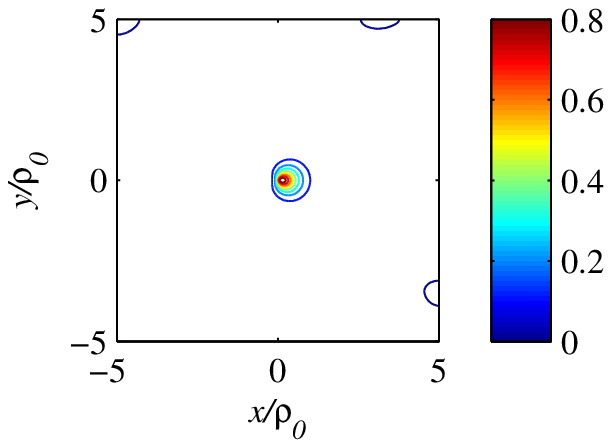}
\label{fig:PatternNoiseR1Time500}
}
\caption{Contours of $\eta$ when thermal noise is present for a system with a dislocation for the case $r_{0}=1$, $u_{1}=-1$ and $v_{0}=1$ at different times.}
\label{fig:PatternDislNoiseR1}
\end{figure}

\begin{figure}[htbp]
\includegraphics{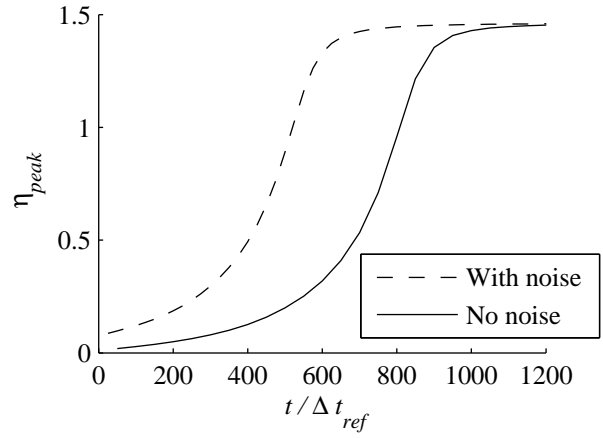} \caption{Comparison of the temporal evolution of $\eta_{peak}$ with and without the presence of thermal noise for $r_{0}=1$, $u_{1}=-1$ and $v_{0}=1$.}
\label{fig:CompNoiseRateTopRpos1Uneg1}
\end{figure}

We now study the influence of the thermal noise term in Eq.~(\ref{eqn:tdgl-eq2}) on the phase transformation kinetics in the presence of a dislocation for one case; $u_{1}=-1$ and $r_{0}=1$, for which results of computations without the noise term were presented earlier, see subsections~\ref{sec:SpatioTempEvol}. Here, the evolution of the order parameter is illustrated in the form of contour plots of $\eta$ at different times, see Fig. \ref{fig:PatternDislNoiseR1}. It is seen that at a very early stage of phase transition a top emerges near the origin in the half-plane where $x>0$, although the variance of the background noise is of the same order of magnitude. With increasing time the thermal noise loses its impact. A single top is expected to evolve, which is confined within a circular area on the compressive side of the dislocation, as in the case with no noise added, cf. Fig.~\ref{fig:SurfsRpos1Uneg1}.  However, as can be seen from Fig.~\ref{fig:CompNoiseRateTopRpos1Uneg1}, there is a difference in the evolution rate if noise is included in the calculations. For the disturbed system, $\eta_{peak}$ evolves faster, in the sense that the growth of the top is larger in the very beginning of the evolution. At later times, the growth rate is very similar between the two cases as can be seen from Fig. \ref{fig:CompNoiseRateTopRpos1Uneg1}. The two curves will coincide if the dashed line is shifted horizontally by about $300\Delta t_{ref}$. But if the variance of the Gausssian distribution is reduced, this shift is also reduced. We should also note that the nucleation rate of the second phase on dislocation is much higher than in homogeneous (defect free solid) nucleation.  This is because the nucleation barrier energy at an edge dislocation is lower than that for homogeneous nucleation; see e.g.  \cite{Gomez_Ramirez_1973}. A similar remark also applies to the growth of second phase \cite{Massih_2009}.

It can be concluded that the results in this study will not differ much if Langevin thermal noise is included in the governing equation or not. The only significant difference that has been noted is that the nucleation of the ordered phase at a dislocation starts earlier if thermal noise is taken into account at $T>T_{c0}$, i.e.  an edge  dislocation is hosted in the solid solution. This can be interpreted as that the noise supports the nucleation at a singularity, and may not suppress it. The main features of the evolution of all the studied cases in this paper will still be effective in the presence of thermal noise, at least with the parameter set and noise distribution used here.


\section{Analytical solutions: Late time growth}
\label{sec:analytic}



It is worthwhile to discuss now the late time growth behavior of the second-phase/matrix interface within the GLK theory. To this end, we consider the equal-time order parameter correlation function
\begin{equation}
\mathcal{C}(\boldsymbol{\rho},t)=\langle\eta(\boldsymbol{\rho},t)\eta(0,t)\rangle.
\label{eqn:correlfun}
\end{equation}
In late times after a deep temperature quench into an unstable state, $\mathcal{C}(\boldsymbol{\rho},t)$ takes a  form \cite{Mazenko_2008},
\begin{equation}
\mathcal{C}(\boldsymbol{\rho},t)=\eta_{0}^{2}\mathit{F}(\boldsymbol{\rho}/L(t)),
\label{eqn:universalfun}
\end{equation}
where $\eta_{0}\equiv\eta_{+}$ is the magnitude of long-time equilibrium value of $\eta(\boldsymbol{\rho},t)$ and $\mathit{F}(\bullet)$ is a universal scaling function, which depends on the symmetry of the system (here a scalar order parameter) and the spatial dimension. Here, $L(t)$ is a characteristic length scale featuring the growth law for the second phase.

Mazenko's  functional-integral method for phase ordering kinetics \cite{Mazenko_1989,Mazenko_1990} shows that the scaling function $\mathit{F}(\bullet)$ obeys the eigenvalue problem
\begin{equation}
-\mu\textbf{x}\cdot\nabla_{x}F=\tan(\pi F/2)+\nabla_{x}^{2}F,
\label{eqn:mazenko-evp }
\end{equation}
where $\textbf{x}=\boldsymbol{\rho}/L(t)$ and $\mu$ is an eigenvalue to be determined. An exact analytical expression for $L(t)$ valid for all times is not known, however, at late times
\begin{equation}
L(t)=\big(4\Gamma_{a}\mu t\big)^{1/2}.
\label{eqn:lca-law}
\end{equation}
Equation (\ref{eqn:lca-law}) is the manifest Lifshitz \cite{Lifshitz_1962}, Cahn and Allen \cite{Cahn_Allen_1977} curvature-driven growth law for non-conserved systems at late times. Mazenko \cite{Mazenko_1990} has determined numerically the eigenvalue $\mu=\mu^{*}(d)$, which is space dimension dependent, namely, $\mu=1.104$ and $\mu=0.5917$ for $d=2$ and $d=3$, respectively.

\begin{figure}[htbp]
\includegraphics[width=0.6\textwidth]{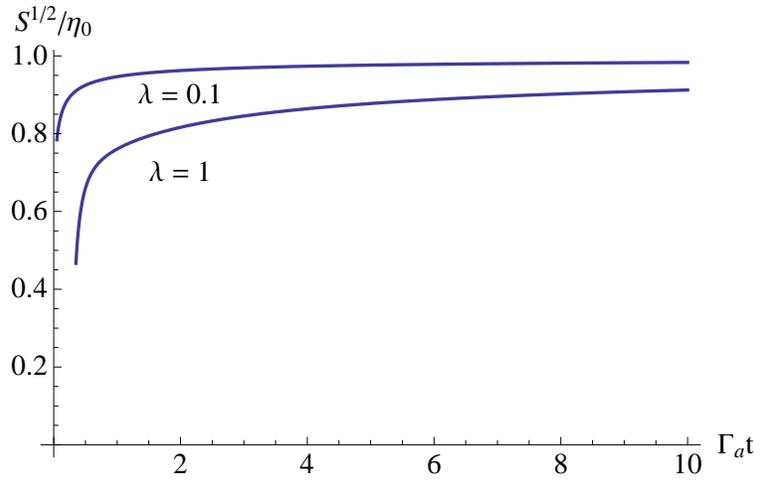}
\caption{Late temporal evolution of the root mean square of the order parameter scaled with the equilibrium value $\eta_{0}$ at two values of the coefficient $\lambda$.}
\label{fig:msq-sum}
\end{figure}

Let us now using Mazenko's method calculate the temporal evolution of the ensemble average of the order parameter $\langle\eta^{2}(\boldsymbol{\rho},t)\rangle^{1/2}$, cf. Eq. (\ref{eqn:meansquare}), at the interface for $r_{1}<0$ and $u_{1}<0$. The computations outlined in Appendix \ref{sec:Mazenko} give
\begin{equation}
\mathcal{S}(t)=\eta_{0}^{2}\Big(1+\sum_{n=0}^{\infty}\frac{(-1)^{2n+1}a_{2n+1}}{L(t)^{2n+1}}\Big),\label{eqn:msq-infsum}
\end{equation}
where $a_{k}$'s are constants for a given state of the system. This series is asymptotically convergent. In Appendix \ref{sec:Mazenko}, we have given the first three coefficients of the series (\ref{eqn:msq-infsum}), which are adequate for convergence. Figure \ref{fig:msq-sum} depicts $\sqrt{\mathcal{S}}/\eta_{0}$ as a function of the dimensionless time for two values of $\lambda\equiv|r_{1}|v_{0}/u_{1}^{2}$ using Eqs. (\ref{eqn:msq-infsum}) with up to $n=3$, and (\ref{eqn:lca-law}) with $d=2$. It is seen that an increase in $\lambda$ slows down the completion of equilibrium. Table \ref{tab:msq-sum} shows this quantitatively.

\begin{table*}
\caption{\label{tab:msq-sum}Computational results of late time temporal evolution of the mean order parameter using Eqs. (\ref{eqn:msq-infsum}) with up to $n=3$, and (\ref{eqn:lca-law}) with $d=2$, cf. Fig. \ref{fig:msq-sum}.}
\begin{ruledtabular}%
\begin{tabular}{ccc}
$\Gamma_{a}t$  & $\sqrt{\mathcal{S}}/\eta_{0}\,(\lambda=0.1)$  & $\sqrt{\mathcal{S}}/\eta_{0}\,(\lambda=1)$ \tabularnewline
\hline
10  & 0.983266  & 0.912301\tabularnewline
100  & 0.994734  & 0.972326\tabularnewline
1000  & 0.998337  & 0.991307\tabularnewline
10000  & 0.999475  & 0.997258\tabularnewline
\end{tabular}\end{ruledtabular}
\end{table*}

It may be of interest to compare the results of the late time evolution of $\langle\eta^{2}(\boldsymbol{\rho},t)\rangle^{1/2}$ with the numerical computations of $\eta_{mean}$ made on the entire range of $t$ depicted in Fig. \ref{fig:CompNoiseMeanEtaVsTime} and discussed in Sec. \ref{sec:NoiseNoDisl} for dislocation-free solid. As mentioned there, the curves in Fig. \ref{fig:CompNoiseMeanEtaVsTime} fit the logistic curve, given by  Eq. (\ref{eq:logisic-fit}), fairly well.  Although, we do not know where exactly to  draw the line for late time growth and even the two quantities may not directly be comparable, we have attempted to fit the calculation output for $\lambda=1$, Fig. \ref{fig:msq-sum}, to a logistic curve,  Eq. (\ref{eq:logisic-fit}), in the time interval $\Gamma_at \in [0.5,10]$. The fit was adequate with $a=0.902,b=0.386$, $c=0.386$, and  $t \to \Gamma_at$ for $\mathcal{S}^{1/2}/\eta_{0}\equiv \langle\eta^{2}(\boldsymbol{\rho},t)\rangle^{1/2}/\eta_0$. This shows that our numerical solutions for $\eta_{mean}$ follow the same evolution trend as the analytical solution using Mazenko's theory.


\section{Summary and Conclusions}
\label{sec:Conclude}

In this work, we have used the time-dependent Ginzburg-Landau (TDGL) equation for a single component non-conservative structural order parameter $\eta$ to model the spatio-temporal evolution of a second phase in the vicinity of an edge dislocation in an elastic crystalline solid. A symmetric Landau potential of a sixth-order, $\eta^6$, was employed. The phase-field equation was solved numerically using a finite volume method, where a wide range of parameter sets is explored. Computations were performed for the situations where the temperature was held above the transition temperature of a defect-free crystal, $T_{c0}$, as-well-as below it. In both cases, the influence of the elastic properties of the material and the strength of interaction between the order parameter and the elastic displacement field were examined by varying a model parameter that comprises these properties.

We found that the introduction of a dislocation always triggers nucleation of a second phase. The phase transition initiates in the vicinity of the dislocation line in the region where stresses are compressive, regardless of the parameter setting. If the temperature is above $T_{c0}$ and the elastic interaction is moderate, we found that the second phase grows locally within a confined space. However, if the elasticity and/or elastic interaction is large, eventually the phase transition will spread throughout the whole material even though the temperature exceeds $T_{c0}$. In the regions where the stresses induced by the dislocation are tensile in character, the phase transition is suppressed, though not fully. If the system is quenched below $T_{c0}$, the entire material transforms and steady state is first reached on the compressive side of the dislocation. On the side with tensile stresses, the evolution of ordering is held back. The steady-state distribution of phases is estimated by considering a modified Landau type potential.

The influence of the Langevin thermal noise term in the TDGL equation was also examined. We found that if the dislocation is introduced in the crystal above $T_{c0}$, the thermal noise will support the nucleation of the ordered second phase, and steady state will be reached earlier than if the thermal noise were absent. But if a dislocation is introduced in a solid solution whilst the system is being quenched below $T_{c0}$, the evolution of second phase ordering will not be affected significantly by the noise term.

For a dislocation-free solid, we compared our numerical computations for a mean-field (spatially-averaged) order parameter as a function of time with the late time growth of the ensemble-averaged order parameter calculated from Mazenko's theory of domain growth, and found that both results follow late time logistic curves.

The present work is part of a larger study on how singular stress fields affect phase transformations in crystalline materials, where both dislocations and cracks are considered. The subsequent steps of this study aim to extend the present calculations to include a two-component field structural order parameter, and to consider the coupling of the composition, obeying a conserved kinetic equation, and the two-component order parameter. This deliberation not only will describe the formation and growth of second phase in the vicinity of defects but also will tell its orientation in the presence of applies stress.


\begin{acknowledgments}
The work was supported by the Knowledge Foundation of Sweden grant number 2008/0503 and 2011/0215.
\end{acknowledgments}


\bibliography{micro}

\clearpage{}

\appendix


\section{Evolution of the interface in quenched phase}
\label{sec:Mazenko}

Here we outline the method introduced by Mazenko and coworkers \cite{Mazenko_1989,Mazenko_1990,Liu_Mazenko_1992} to study the growth kinetics of quenched systems from early through late times which comprised sharp interfaces in the late stage development. We tailor the approach to our application, i.e. the growth of a second phase within the framework of non-conserved $\eta^{6}$ Ginzburg-Landau theory.

We consider the time-dependent Ginzburg-Landau equation for an elastic body containing a dislocation. As in Sec. \ref{sec:Noise}, the considered equation is supplied by an initial probability distribution governing $\eta$ at time $t=0$ in the solid solution $T>T_{c}$, which is supposed to be Gaussian with the initial correlation $\langle\eta(\boldsymbol{\rho},0)\eta(\boldsymbol{\rho}\rq{},0)\rangle=2k_{B}T\delta(\boldsymbol{\rho}-\boldsymbol{\rho}\rq{})$
and $\langle\eta(\boldsymbol{\rho},0)\rangle=0$. The system is then quenched to a state below $T_{c}$ upon which a second phase is formed around the dislocation. One can assume that the system is quenched to zero temperature so the thermal noise can be set equal to zero for $T<T_{c}$.

A key notion in Mazenko's theory is the separation of the order parameter field $\eta$ into a peak contribution $\sigma$ and a fluctuating term $\delta\eta$, viz.
\begin{equation}
\eta(\boldsymbol{\rho},t)=\sigma[s(\boldsymbol{\rho},t)]+\delta\eta(\boldsymbol{\rho},t),
\label{eq:MazAnsatz}
\end{equation}
where $\sigma[s(\boldsymbol{\rho},t)]$ is a functional of the auxiliary field $s(\boldsymbol{\rho},t)$ and $s$ is given a physical interpretation, near interfaces (walls), as a coordinate normal to the wall. So $\sigma[s]$ may be considered as the equilibrium interfacial profile depending on $s$. Furthermore, it is assumed that $s$ is proportional to the characteristic length $L$ that scales the system. More precisely, one defines
\begin{equation}
\mathfrak{s}^{2}\equiv\langle s^{2}\boldsymbol{(\rho},t)\rangle=L^{2}/\pi,
\label{eqn:charlength}
\end{equation}
where the angular brackets denote the ensemble average over random
initial conditions, and also one posits that $\mathfrak{s}$ and $L$
are increasing function of time.

Ignoring now the fluctuating term in Eq. (\ref{eq:MazAnsatz}), then placing it into the expression for the free energy functional, Eq. (\ref{eqn:toten1}), and using the Euler-Lagrange equation, we obtain
\begin{equation}
\frac{g_{a}}{2}\frac{\mathrm{d}^{2}\sigma}{\mathrm{d}s^{2}}-\big(-|r_{1}|\sigma+u_{1}\sigma^{3}+v_{0}\sigma^{5}\big)=0,
\label{eqn:gle-1}
\end{equation}
\noindent which is precisely the time-independent Ginzburg-Landau equation near an isolated defect, cf. Eq. (\ref{eqn:tdgl-eq2}). We can rewrite this last equation in terms of scaled (dimensionless) variables in the form
\begin{equation}
\frac{1}{2}\frac{\mathrm{d}^{2}\tilde{\sigma}}{\mathrm{d}\hat{s}^{2}}+\bar{\sigma}-\mathrm{sgn}(u_{1})\bar{\sigma}^{3}-\lambda\bar{\sigma}^{5}=0,\label{eqn:gle-2}
\end{equation}
where $\bar{\sigma}=\sigma/\sigma^{\ast}$, $\sigma^{\ast}=(|r_{1}|/|u_{1}|)^{1/2}$, $\hat{s}=(r_{1}|/g_{a})^{1/2}s$, and $\lambda=|r_{1}|v_{0}/u_{1}^{2}$. We now, for convenience, drop the bar and the hat symbols from the variables and rewrite Eq. (\ref{eqn:gle-2}) in the form
\begin{eqnarray}
\frac{1}{2}\frac{\mathrm{d}^{2}\sigma}{\mathrm{d}s^{2}} & = & V^{\prime}[\sigma],
\label{eqn:gle-3}\\
V^{\prime}[\sigma] & = & -\sigma+\mathrm{sgn}(u_{1})\sigma^{3}+\lambda\sigma^{5}.
\end{eqnarray}
\noindent Moreover, we impose the boundary condition
\begin{equation}
\lim_{s\rightarrow\infty}\frac{\mathrm{d}\sigma}{\mathrm{d}s}=0.
\label{eqn:gle-bc}
\end{equation}
Integration of (\ref{eqn:gle-3}) gives
\begin{equation}
s=\pm\frac{1}{2}\int_{0}^{\sigma}\frac{\mathrm{d}x}{\sqrt{V[x]-V[\eta_{0}]}}.
\label{eqn:gle-sol1}
\end{equation}
where the critical points $\pm\eta_{0}^{2}$ are found from $V^{\prime}[\eta_{0}]=0$,
\begin{equation}
\eta_{0}^{2}=\frac{1}{2\lambda}\Big(-\mathrm{sgn}(u_{1})+\sqrt{1+4\lambda}\Big).
\label{eqn:gle-turp}
\end{equation}
Furthermore,
\begin{equation}
V[\eta_{0}]=-\frac{1}{2}\eta_{0}^{2}\Big(1-\mathrm{sgn}(u_{1})\eta_{0}^{2}\Big)+\frac{\lambda}{6}\eta_{0}^{6}.
\label{eqn:gle-pot}
\end{equation}
We evaluate the integral in (\ref{eqn:gle-sol1}) to obtain
\begin{eqnarray}
s & = & \pm\xi_{0}\Big(\arctan\Theta+i\frac{\pi}{2}\Big),\label{eqn:gle-sol2}\\
\text{with}\quad\xi_{0} & = & (2-\mathrm{sgn}(u_{1})\eta_{0}^{2})^{-1/2},\\
\Theta^{2} & = & \frac{2}{3}\xi_{0}^{2}\Big(\eta_{0}^{2}+1+\frac{1}{2\sigma^{2}}(\eta_{0}^{2}+4)\Big).\label{eqn:Theta}
\end{eqnarray}
Inverting Eq. (\ref{eqn:gle-sol2}) to find $\sigma$ as function
of $s$, after some manipulation, we obtain
\begin{eqnarray}
\sigma & = & \frac{\eta_{0}\sqrt{1-\epsilon}\tanh\tilde{s}}{\sqrt{1-\epsilon\tanh^{2}\tilde{s}}},\label{eqn:sig-v-s}\\
\text{with}\quad\epsilon & = & \frac{2}{3}\Big(\frac{\eta_{0}^{2}+1}{\eta_{0}^{2}+2}\Big)\quad\text{for}\quad u_{1}<0,\\
\tilde{s} & = & s/\xi_{0}.
\end{eqnarray}
Equation (\ref{eqn:sig-v-s}) is identical to Mazenko's result \cite{Mazenko_1990} for $u_{0}>0$ except for some modifications in the definitions of $\epsilon$ and $\xi_{0}$.

It can be shown that as $\lambda$ is increased $\sigma$ tends toward the origin. Moreover, for large $s$, $\sigma\sim\eta_{0}$. This trend can seen by expanding $V[x]$ near $\eta_{0}$ and evaluating the integral in Eq. (\ref{eqn:gle-sol1}) to obtain
\begin{equation}
\sigma[s]=\pm\eta_{0}\Big(1-\exp[-\sqrt{2V^{\prime\prime}(\eta_{0})}\; s]\Big).
\label{eqn:gle-asympt}
\end{equation}
We should also recall that for $u_{1}>0$ the case $\lambda=0$ gives $\eta_{0}=\pm1$ with $\sigma=\pm\tanh[s]$.

One quantity of interest for our analysis is the interfacial width $\xi$ \cite{Mazenko_1990} defined by
\begin{equation}
\xi=\frac{1}{\eta_{0}^{2}}\int_{-\infty}^{+\infty}\big[\eta_{0}^{2}-\sigma^{2}[x]\big]\mathrm{d}x.\label{eqn:width-def}
\end{equation}
Substituting for $\sigma(x)$ in the integrand from Eq. (\ref{eqn:sig-v-s}) and evaluating the integral gives
\begin{equation}
\xi=2\frac{\xi_{0}}{\sqrt{\epsilon}}\mathrm{arctanh}[\sqrt{\epsilon}\,].
\label{eqn:width-sol1}
\end{equation}
Using a standard identity in hyperbolic functions, we write this formula as
\begin{equation}
\xi=\frac{\xi_{0}}{\sqrt{\epsilon}}\log{\frac{1+\sqrt{\epsilon}}{1-\sqrt{\epsilon}}},\quad\mathrm{for}\quad0\le\epsilon<1.
\label{eqn:width-sol2}
\end{equation}
Let us next calculate the evolution of the local order parameter, cf. Eq. (\ref{eqn:meansquare}), in the $s$-coordinate
\begin{equation}
\mathcal{S}(t)=\langle\sigma^{2}[s]\rangle,
\label{eqn:msq-1}
\end{equation}
with
\begin{equation}
\langle\sigma^{2}[x]\rangle=\int_{-\infty}^{+\infty}\sigma^{2}[x]\mathcal{P}(x)\mathrm{d}x,\label{eqn:msq-def}
\end{equation}
where $\mathcal{P}(x)$ is taken to be a Gaussian distribution of a real random variable $x$ with a variance $\mathfrak{s}$, defined in Eq. (\ref{eqn:charlength}), given by
\begin{equation}
\mathcal{P}(x)=\frac{1}{\sqrt{2\pi\mathfrak{s}^{2}}}\exp[-x^{2}/2\mathfrak{s}^{2}],
\label{eqn:gdf}
\end{equation}
Again following the procedure delineated in \cite{Mazenko_1990}, we split
\begin{equation}
\sigma^{2}[s]=\sigma^{2}[\infty]+\Delta\sigma^{2}[s],
\label{eqn:sigma-split}
\end{equation}
where $\Delta\sigma^{2}[s]=\sigma^{2}[s]-\eta_{0}^{2}$. Hence Eq. (\ref{eqn:msq-1}) is expressed in the form
\begin{equation}
\mathcal{S}(t)=\sigma^{2}[\infty]+\frac{1}{\sqrt{2\pi\mathfrak{s}^{2}}}\int_{-\infty}^{+\infty}e^{-x^{2}/2\mathfrak{s}^{2}}\Delta\sigma^{2}[x]\mathrm{d}x,
\label{eqn:msq-2}
\end{equation}
Expanding the exponential term in Eq. (\ref{eqn:msq-2}) in infinite series,
\begin{equation}
\mathcal{S}(t)=\eta_{0}^{2}+\frac{1}{\sqrt{2\pi\mathfrak{s}^{2}}}\sum_{n=0}^{\infty}\Big(\frac{-1}{2\mathfrak{s}^{2}}\Big)^{n}\frac{1}{n!}\int_{-\infty}^{+\infty}x^{2n}e^{-x^{2}/2\mathfrak{s}^{2}}\Delta\sigma^{2}[x]\mathrm{d}x,
\label{eqn:msq-exp}
\end{equation}
\noindent where we put $\sigma^{2}[\infty]=\eta_{0}^{2}$ by (\ref{eqn:gle-asympt}). Considering the first three terms in the expansion, then performing the integrations and replacing $\mathfrak{s}$ by $L(t)$ via Eq. (\ref{eqn:charlength}), we obtain
\begin{equation}
\mathcal{S}(t)=\eta_{0}^{2}\Big(1-\frac{a_{1}}{L}+\frac{a_{3}}{L^{3}}-\frac{a_{5}}{L^{5}}+\dots\Big),\label{eqn:msq-bsum}
\end{equation}
where $a_{1}=\xi/\sqrt{2}$, and
\begin{eqnarray}
a_{3} & = & \frac{\pi\epsilon}{2^{3/2}}\frac{\xi}{6}\Big(\xi^{2}+\frac{\pi^{2}\xi_{0}^{2}}{\epsilon}\Big),\label{eqn:msq-coeffs}\\
a_{5} & = & \frac{\pi^{2}\epsilon^{2}}{2^{5/2}}\frac{\xi}{80}\Big(\xi^{4}+\frac{10\pi^{2}}{3\epsilon}\xi_{0}^{2}\xi^{2}+\frac{7\pi^{4}\xi_{0}^{4}}{3\epsilon^{2}}\Big).
\end{eqnarray}
The series (\ref{eqn:msq-bsum}) is asymptotically convergent; the expansion up to $\mathcal{O}(L^{-5})$ provides sufficient accuracy in computations. Hence, if knowing the time-dependence of the characteristic length $L$, then the time evolution of the averaged order parameter can be calcuated from Eq. (\ref{eqn:msq-bsum}).

Cardy \cite{Cardy_1992} has provided an alternative approach to compute the temporal evolution of the non-conserved order parameter in the case of a quench to a temperature at, or just above, the critical temperature, when the slowness in the dynamics is a result of
the critical slowing down of local fluctuations. He found, using the renormalization group and $\epsilon$-expansion technique, to all orders in $\epsilon$, that the local fluctuations in the order parameter scale like $t^{-1/2}$, and have a universal distribution.

\clearpage


\section{Figures}

\clearpage{}

\clearpage{}

\clearpage{}

\clearpage{}

\clearpage{}

\clearpage

\clearpage{}

\clearpage{}

\clearpage{}

\clearpage{}

\clearpage{}

\clearpage{}

\clearpage{}

\end{document}